\documentclass[11pt]{amsart}
\usepackage[english]{babel}
\usepackage{amsmath,amssymb,eucal,graphicx}
\usepackage{amsthm}
\usepackage{hyperref}
\usepackage{amsaddr}

\usepackage[square,comma,numbers,sort&compress]{natbib}
\usepackage[all]{xy}
\usepackage{color}

\begin{document}
\title[Multi-type critical process]{Error-induced extinction in a multi-type critical birth-death process}

\author{Meritxell Brunet Guasch$^1$,\\ P. L. Krapivsky$^2$, Tibor Antal$^1$}
\address{1, School of Mathematics and Maxwell Institute for Mathematical Sciences, University of Edinburgh, Edinburgh EH9 3FD, UK}
\address{2, Department of Physics, Boston University, Boston, Massachusetts 02215, USA;
Santa Fe Institute, Santa Fe, New Mexico 87501, USA
}

\date{\today}

\begin{abstract}  Extreme mutation rates in microbes and cancer cells can result in error-induced extinction (EEX), where every descendant cell eventually acquires a lethal mutation. In this work, we investigate critical birth-death processes with $n$ distinct types as a birth-death model of EEX in a growing population. Each type-$i$ cell divides independently $(i)\to(i)+(i)$ or mutates $(i)\to(i+1)$ at the same rate. The total number of cells grows exponentially as a Yule process until a cell of type-$n$ appears, which cell type can only die at rate one. This makes the whole process critical and hence after the exponentially growing phase eventually all cells die with probability one.
We present large-time asymptotic results for the general $n$-type critical birth-death process. We find that the mass function of the number of cells of type-$k$ has algebraic and stationary tail $(\text{size})^{-1-\chi_k}$, with $\chi_k=2^{1-k}$, for $k=2,\dots,n$, in sharp contrast to the exponential tail of the first type. The same exponents describe the tail of the asymptotic survival probability $(\text{time})^{-\chi_n}$. We present applications of the results for studying extinction due to intolerable mutation rates in biological populations.
\end{abstract}

\maketitle
\section{Introduction}

Genetic defects in DNA replication fidelity and repair result in mutator phenotypes that accelerate the adaptation of microbes and cancer cells. However, certain combinations of mutator alleles increase mutation rates to intolerable levels, resulting in error-induced extinction (EEX), where every cell eventually acquires a lethal mutation \cite{morrison1993pathway,fijalkowska1996mutants}.
Error-induced extinction occurs within a few generations in bacteria and haploid yeast with mutations on both DNA proofreading and repair  \cite{morrison1993pathway,fijalkowska1996mutants, herr2011mutator}. Tumours with mutator phenotypes reach an upper limit of mutations, which has been interpreted as evidence for a maximal mutation rate in cancer \cite{fox2010lethal, schumacher2019cancer}.

Deﬁning the maximum mutation rate is important for understanding the long-term ﬁtness of mutator alleles, with applications for studying the evolution of hyper-mutated tumours and the synthetic lethality of mutator alleles \cite{topatana2020advances}. Experimentally, this amounts to producing cell lines with mutator mutations, and measuring the mutation rates when colonies are viable. This strategy has identified intolerable mutation rates for bacteria \cite{morrison1993pathway}, yeast  \cite{fijalkowska1996mutants, soriano2021expression} and cancer in mice \cite{albertson2009dna}. However, beyond a threshold, the cell lines are not viable or acquire antimutator alleles. Thus, it is not possible to study the behaviour of populations at the limiting mutation rate and the evolutionary process underlying their extinction. Open questions include: for how many generations, and how large can populations grow at the limit mutation rate? What is the genetic structure of populations undergoing EEX? Can mutation drive the extinction of an exponentially growing population? 

Here we propose a continuous-time multi-type critical birth-death process that allows theoretical exploration of the questions above, amongst other applications of biological interest. This models a population of cells that divide, die or mutate independent of each other, where the time between events is exponentially distributed. To mimic the behaviour of cells at the limiting mutation rate, we set the rates of division $(\alpha)$ and death $(\beta)$ or mutation $(\nu)$ to be balanced $(\alpha=\beta + \nu)$.
In the simplest model for EEX, there is a single type of cells that divides $(1)\to(1)+(1)$ or acquires a lethal mutation  $(1)\to \emptyset$  at the same rate.
This is the case of haploid yeast and bacteria with mutation rates of one lethal mutation per cell division, which undergo EEX within a few generations \cite{morrison1993pathway,fijalkowska1996mutants, herr2011mutator}. A more biologically interesting case is to allow multiple mutations to accumulate before a lethal mutation arrives. We model this by considering different cell types, where we call a cell type-$i$ if it has accumulated $i$ mutations. We assume that there is a maximal number $n$ of mutations a cell can bear. Note that in this model mutations accumulate consecutively, therefore the lethal mutation is the $n$th mutation, not \emph{a} particular mutation. 

\begin{figure}[h!]
\centering
\includegraphics[scale=0.6]{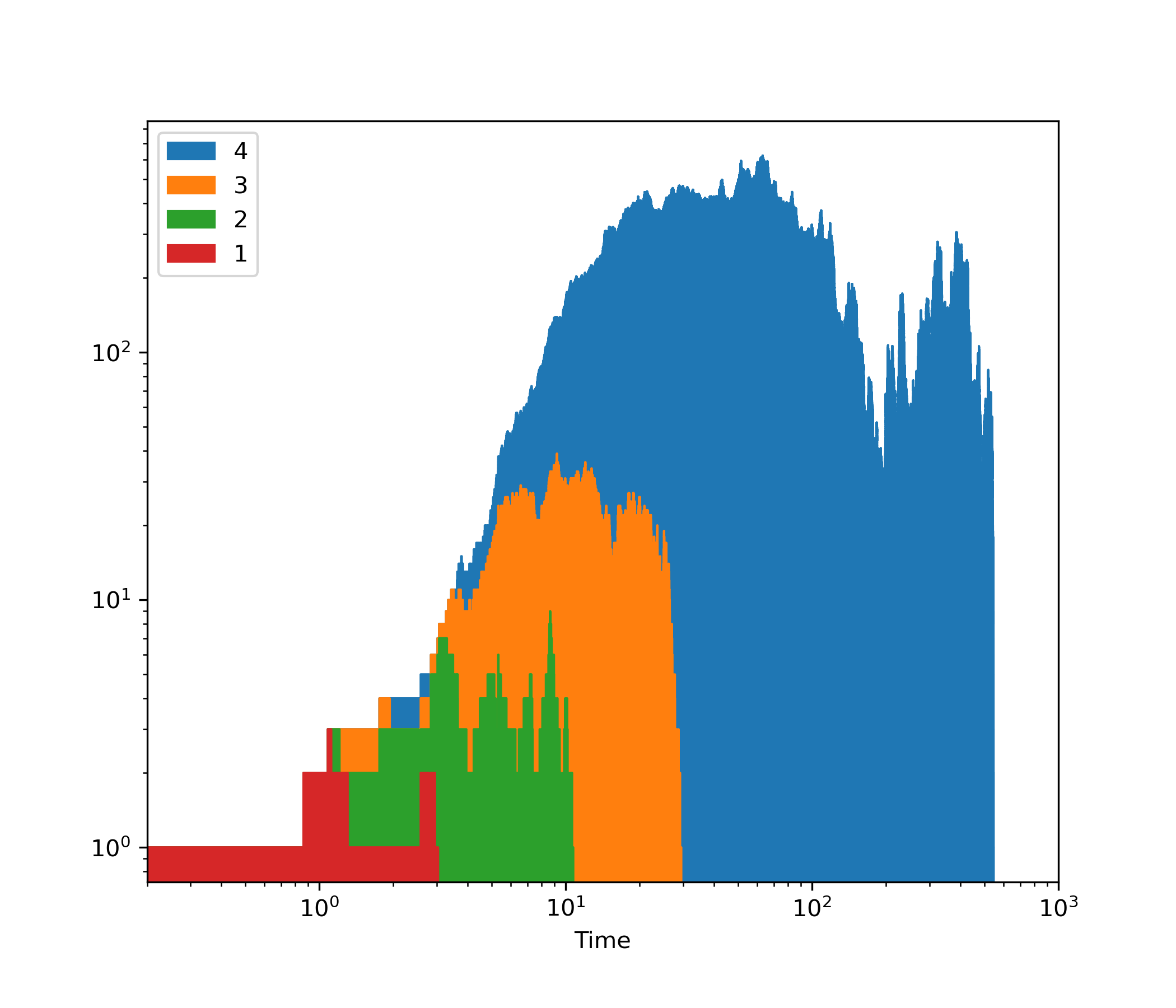}
\caption{An example simulation of the process of \eqref{fig:bd_multi} run until extinction with n=4 types, where each coloured area represents the number of cells of a type, with different types piled on top of each other. Hence the envelope is the total population size. A single initial type-1 cell resulted in a short-lived type-1 population which seeded a type-2 population before its extinction. Note how type-3 and type-4 cells were initiated multiple times. One can observe the initial fast growth of the total population before extinction.}\label{fig:1}
\end{figure}

In the simplest multi-type critical process, each type-$i$ cell divides independently $(i)\to(i)+(i)$ at rate one, or accumulates a new mutation $(i)\to(i+1)$ also at rate one, except type-$n$ cells, which die at rate one. We represent this birth-mutation process by the following scheme
\begin{equation}
\begin{split}
\xymatrix{(1)+(1)           & (2)+(2)          &                    & (n)+(n)        &\\
                (1) \ar[u] \ar[r] & (2)\ar[u] \ar[r] & \cdots \ar[r]  & (n) \ar[u] \ar[r] &\emptyset}
\end{split} \label{fig:bd_multi}   
\end{equation}
A more general version comes from introducing death and allowing different types of cells to divide and mutate at different rates. We refer to this as the $n$-type birth-death process, which can be illustrated as
\begin{equation}
\begin{split}
\xymatrix{(1)+(1)			& (2)+(2)				&			& (n)+(n)		\\
                (1) \ar[u]_{\alpha_1} \ar[r]_{\nu_1} \ar[d]^{\alpha_1-\nu_1}	& (2)\ar[u]_{\alpha_2} \ar[r]_{\nu_2} \ar[d]^{\alpha_2-\nu_2} 	& \cdots \ar[r]_{\nu_{n-1}} 	& (n) \ar[u]_{\alpha_n} \ar[d]^{\alpha_n}\\
                \emptyset			& \emptyset			& 			& \emptyset }
\end{split}                
\label{fig:bd_multid}
\end{equation}
Note that for any given type, cells appear (via division) and disappear (via mutation or death) at the same rate, and thus all types remain critical. The overall process is critical, hence eventually all cells go extinct with probability one. This can be seen intuitively, since critical type-1 cells go extinct without a source, at which point the same can be said about type-2 cells, and so on. Figure \ref{fig:1} shows the evolution of the number of cells of each type in a four-type process of \eqref{fig:bd_multi}. Note that each type grows faster than the previous and takes longer to become extinct. Thus, the more mutations that can accumulate before a lethal mutation, the longer it takes for the population to undergo EEX. Interestingly, the population grows approximately exponentially during a transient, but eventually goes extinct with probability one. This behaviour cannot be captured by a supercritical process with an accumulation of deleterious mutations. In that case, if the initial types of cells have higher fitness, these will overtake any subsequent less fit types, resulting in a positive probability of survival of the population. Moreover, the $n$-type critical process does not require a reduction in fitness until the last lethal mutation, naturally capturing the phenomenon of EEX.

Due to their applicability to model exponentially growing populations, multi-type, super-critical processes with consecutive mutations (decomposable) have been extensively studied \cite{Athreya:2004, kesten67, Durrett:2015, nicholson2022mutation}. In the critical case, the asymptotic behaviour has been studied by many authors \cite{sevast1959transient, Chistyakov59, mullikin1963limiting}. Foster and Ney \cite{Ney74} derived the asymptotic survival probability for discrete-time models. Similar results were obtained simultaneously by Ogura \cite{ogura1975asymptotic}, also including continuous time processes. Later on, Foster and Ney \cite{foster1978limit} proposed limit theorems for the generating functions of population sizes, conditioned on the survival of the first type of cells, a condition less relevant for biological applications.

In this work, we derive large-time asymptotic solutions of multi-type critical birth-death processes exploiting the fact that, after an initial stage of exponential growth, the population is dominated by the last type. More precisely, we show that, for a large time, the system is non-empty with probability proportional to $(\text{time})^{-\chi_n}$, where $\chi_n=2^{1-n}$, establishing the relationship between the time-dependent survival probability and the maximal number of mutations that can accumulate. The exponents of the survival probability were derived by \cite{Ney74, ogura1975asymptotic} through a different approach. Knowing the asymptotic survival probability allows us to find an appropriate scaling of the system, and derive the limiting distribution for the number of cells of a given type present at time $t$. We find asymptotic solutions for the distribution of cells of type $k=1,\dots,n$, as well as the total number distribution. Our methods extend the exact solution and limit results \cite{Antal:2011} for the two-type critical birth-death case, and complement the results of recent work on super-critical processes  \cite{nicholson2022mutation}. Remarkably, we find that the distributions for the number of cells of a given type and the total number distributions have algebraic and stationary tails, described by the same exponents $\chi_n$ as the survival probabilities. This provides interesting biological insight into the behaviour of the modeled populations, showing that they can reach a stationary growth phase before extinction. We derive further estimates of interest for studying cancer and bacteria growth, including the distributions of time of arrival and extinction of cells with an arbitrary number of mutations.

The next sections progressively build up to the main results of this work, which can be found in Section~\ref{multi}. We   introduce the simplest critical processes in sections~\ref{sec:one-type} and ~\ref{inf-type}. We present exact solutions to the two-type critical process in Section~\ref{two} and derive asymptotic solutions to the $n$-type case in Section~\ref{multi}. In Appendix~\ref{app:ndeath}, we examine the more general processes including death and arbitrary birth and mutation rates, and show that the behavior is essentially the same as in the simplest critical birth-death process. Hence in the bulk of the paper, we limit ourselves to the simplest version.  Example applications of the results for studying EEX in microbes and tumours, as well as connections to experimental work, can be found in Section~\ref{sec:exs}.

\section{Single type}
\label{sec:one-type}
In the simplest case, EEX occurs because cells acquire a lethal mutation at the same rate as cell division \cite{morrison1993pathway,fijalkowska1996mutants, herr2011mutator}. This is represented by a single type critical process. Let us recall its basic properties \cite{Athreya:2004}. We denote by $Z_i(t)$ the number of type-$i$ cells at time $t$. We start with a single type-1 cell and study $Z_1(t)$ via the generating function 
$$
\mathcal{Z}_{1,1}(x_1,t)=\mathbb{E}(x_1^{Z_1(t)}|Z_1(0)=1) 
= \sum_{a\ge 0} P_a(t) x_1^a,
$$ 
where the first index of $\mathcal{Z}_{1,1}$ refers to the type of the initial cell while the second to the number of types considered, and $P_a(t)=\mathbb{P}(Z_1(t)=a| Z_1(0)=1)$ is the probability of having $a$ cells at time $t$. This generating function satisfies the backward Kolmogorov equation
$\partial_t \mathcal{Z}_{1,1} = (1-\mathcal{Z}_{1,1})^2$ with $\mathcal{Z}_{1,1}(x_1,0)=x_1$, from which 
\begin{equation}
\label{onetype_gen}
\mathcal{Z}_{1,1}(x_1,t) = \frac{t(1-x_1)+x_1}{t(1-x_1)+1}.
\end{equation}
By expanding the generating function around $x_1=0$ we obtain the probability to have $a$ cells at time $t$,
\begin{equation}
\label{CBP:Pmt}
P_a(t)
= \frac{1}{a!} \partial_{x_1}^a \mathcal{Z}_{1,1}(0,t)
= \frac{1}{(1+t)^2} \left( \frac{t}{1+t} \right)^{a-1} 
\end{equation}
for $a\ge 1$, and the survival probability 
\begin{equation}
\label{CBP:St}
S_{1,1}(t) = 1- \mathcal{Z}_{1,1}(0,t) = 1 - P_0(t) = \frac{1}{1+t}.
\end{equation}
Hence the number of cells conditioned on survival has a geometric distribution
$$
 Z_1(t)|\{Z_1(t)>0\} \sim \mathrm{Geo}\left(\frac{1}{1+t}\right).
$$
Notice that the average number of cells remains constant, 
\begin{equation*}
\label{m-av}
\mathbb{E}Z_1(t) = \partial_{x_1} \mathcal{Z}_{1,1}(1,t) =  \sum_{a\geq 1}aP_a(t) = 1,
\end{equation*}
throughout the evolution. Although the probability of extinction for $t\to \infty$ is one, it takes a long time, since for $T=\inf\{t:Z_1(t)=0\}$, 
$$
\mathbb{E}T=\int_0^\infty \mathbb{P}(T>t) dt = \int_0^\infty S_{1,1}(t) dt = \infty.
$$

It is well known \cite{Athreya:2004} that conditioned on survival $Z_1(t)/t$ converges in distribution to an exponential random variable 
\begin{equation}
\label{onetype_scaling}
 \frac{Z_1(t)}{t}|\{Z_1(t)>0\}
 \to Y_1 \sim
 \mathrm{Expo}(1)
\end{equation}
for $t\to\infty$. This is immediate from the properties of a geometric distribution when taking the limit $t,a\to\infty$ limit with $a/t=y$ constant
$$
 \mathbb{P}\left(\frac{Z_1(t)}{t}>y|Z_1(t)>0\right) 
 = \left(\frac{t}{t+1}\right)^{ty} \to e^{-y} = \mathbb{P}(Y_1>y).
$$
One can also derive this from the generating function. First note that $\mathbb{P}(Z_1(t)=a|Z_1(t)>0) = P_a(t)/(1-P_0(t))$ for $a\ge 1$, hence the conditional generating function
\begin{equation}
\label{onetypelim}
 \mathbb{E}(x_1^{Z_1(t)}|Z_1(t)>0) = \sum_{a\ge 1} \frac{P_a(t)}{1-P_0(t)} x_1^a 
 = \frac{\mathcal{Z}_{1,1}(x_1,t)-P_0(t)}{1-P_0(t)}.
\end{equation}
The large $a$ limit is related to $x_1\approx 1$ so, in order to get a non-trivial limit, we write $x_1=1-p/t$ with $p$ constant and notice that for $a,t\to\infty$, the right-hand side of \eqref{onetypelim} converges 
$$ 
 \frac{\mathcal{Z}_{1,1}(1-p/t,t)-P_0(t)}{1-P_0(t)}
 = \frac{1-p/t}{1+p} 
 \to \frac{1}{1+p}.
$$
The convergence of \eqref{onetypelim} requires that $tP_a(t)/(1-P_0(t))\sim t^2 P_a(t)\to f_{Y_1}(y)$ in order to get
$$
 \sum_{a\ge 1} \frac{P_a(t)}{1-P_0(t)} x_1^a 
 = \sum_{a\ge 1} t \frac{P_a(t)}{1-P_0(t)} \left(1-\frac{p}{t}\right)^{yt} \frac{1}{t}
 \to \int_0^\infty f_{Y_1}(y) e^{-py} dy = \mathbb{E}e^{-pY_1}
$$
to converge to the Riemann integral, which is the Laplace transform of the density $f_Y(y)$. The two limits are of course the same, hence 
\begin{equation}
\label{onetypelimit}
 \lim_{t\to\infty} 
 \frac{\mathcal{Z}_{1,1}(1-p/t,t)-P_0(t)}{1-P_0(t)}
 =  \mathbb{E} e^{-pY_1} 
 = \frac{1}{1+p}.
\end{equation}
    By inverting \eqref{onetypelimit}, the Laplace transform of the density $f_{Y_1}(y)$, we indeed obtain that $Y_1\sim \mathrm{Expo}(1)$.

\section{Infinite types}
\label{inf-type}

An interesting special case is obtained when we consider infinitely many types ($n=\infty$) in the process described by scheme \eqref{fig:bd_multi}. Biologically, this represents a population at a limiting mutation rate (hence every sub-population goes extinct) but in which infinitely many mutations can accumulate (hence the total population grows exponentially). 

Since $n=\infty$, there is no extinction, but up to type $m$ the process is identical to the $m$-type model. What becomes simple though in the infinite type model is the total number of cells: if we disregard the types of cells then $Z(t) = \sum_{i\ge 1} Z_i(t)$ is a Yule process with rate one \cite{Athreya:2004}. The generating function of a Yule process $\mathcal{Z}(x,t)=\mathbb{E}(x^{Z(t)}|Z(0)=1)$ can be obtained from the backward Kolmogorov equation
$\partial_t \mathcal{Z} = \mathcal{Z}(\mathcal{Z}-1)$ with $\mathcal{Z}(x,0)=x$, which leads to
$$
\mathcal{Z}(x,t) = \frac{x}{x+(1-x)e^t}.
$$
By expanding this around $x=0$, we obtain the mass function of the total number of cells in the infinite type process with a single initial type-1 cell
\begin{equation*}
\Pi_s(t) = \mathbb{P}(Z(t)=s|Z_i(t)=\delta_{i1}) 
= \frac{1}{s!} \partial_x^s \mathcal{Z}(0,t)
= e^{-t}\big(1-e^{-t}\big)^{s-1},
\end{equation*}
The mean number of cells grows exponentially as 
$\mathbb{E}Z(t)=\partial_x \mathcal{Z}(1,t)=e^t.$
For finite multi-type critical branching processes, $\Pi_s(t)$ 
is not known in general. This is addressed in the next sections.

\section{Two types}
\label{two}

Consider now the simplest two-type critical branching process, which models a population in which a maximal of two mutations can accumulate. This could be interpreted as a population of diploid cells with a single locus for a lethal mutation, in which double allelic mutation is required to acquire the lethal phenotype. The two-type critical birth-death process is represented by 
\begin{equation}
\begin{split}
\xymatrix{(1)+(1)                 & (2)+(2)                &\\
                (1) \ar[u] \ar[r] & (2)\ar[u] \ar[r] & \emptyset}
\end{split}                
\label{proc_2type}
\end{equation}
where all steps occur at equal rates; we set these rates to unity for simplicity. The general case with death is considered Appendix~\ref{app:two}. 
While the single type birth-death process is easily soluble, the two-type birth-death process reduces to generally unsolvable Riccati equations, which were recently shown to be solvable for certain birth-death models with general rates \cite{Antal:2010, Antal:2011}. For the critical case, the situation slightly simplifies and the results are more explicit. Although the analytic solution for the two-type critical case was published in \cite{Antal:2011}, here we re-derive the solution in more detail. 
These exact results will be useful for checking the validity of the asymptotic results we derive in Section~\ref{multi} for general $n$, which is the only available approach for $n\ge 3$.

Since the process is decomposable, the generating function for type-$1$ cells is the same as in the single type process given in \eqref{onetype_gen}.
The generating function for both cell types, starting with a single initial type-$i$ cell is
\begin{equation}
\label{Pxy}
 \mathcal{Z}_{i,2}(x_1,x_2,t) = \mathbb{E} (x_1^{Z_1(t)} x_2^{Z_2(t)} | Z_j(t) = \delta_{ij})
\end{equation}
for $i=1, 2$, and hence the initial conditions are
\begin{subequations}
\begin{align}
  \label{in-PA}
  \mathcal{Z}_{1,2}(x_1,x_2, 0) &= x_1\\
  \label{in-PB}
  \mathcal{Z}_{2,2}(x_1,x_2, 0) &= x_2.
\end{align}
\end{subequations}
A convenient setting for studying the two-type branching process is provided by the {\em backward} Kolmogorov equations
\begin{subequations}
\begin{align}
  \label{PA}
 \partial_t \mathcal{Z}_{1,2} &= \mathcal{Z}_{1,2}^2+\mathcal{Z}_{2,2}-2\mathcal{Z}_{1,2}\\
  \label{PB}
 \partial_t \mathcal{Z}_{2,2} &=  \mathcal{Z}_{2,2}^2 + 1 - 2\mathcal{Z}_{2,2}.
\end{align}
\end{subequations}

Let $\sigma$ be an operator that increases the indices of a function by one, 
\begin{equation}
    \sigma(\mathcal{F}_{i,j}(x_i,x_j))=\mathcal{F}_{i+1,j+1}(x_{i+1},x_{j+1}) \label{eq:sigmaop}.
\end{equation} 
Note that in the general case (see Appendix~\ref{app:ndeath}) $\sigma$ increases the indices of the rates $\alpha_i, \chi_i, \nu_i$ too. Since we are considering decomposable processes, one can easily see that $\mathcal{Z}_{i+1,j+1}=\sigma(\mathcal{Z}_{i,j})$, namely the $j$-type process starting with a single $i$ cell is the same as the $j+1$ process starting with a single $i+1$ cell, up to a change in indices. In particular, $\mathcal{Z}_{2,2}=  \sigma(\mathcal{Z}_{1,1})$, where $\mathcal{Z}_{1,1}$ is the generating function of the single type process, given in \eqref{onetype_gen}.

\subsection{Survival probabilities}
\label{sec:2sp}

Before solving Eqs.~\eqref{PA}--\eqref{PB}, we consider the simpler task of finding the survival probability of the system, where `survival' refers to the situation when there are alive cells of any type at time $t$,
$$
 S_{i,2}(t)
 = \mathbb{P}(Z_1(t)+Z_2(t)>0| Z_k(0)=\delta_{i,k}).
$$ 
The survival probabilities are related to the corresponding generating functions for type-2 cells, $S_{1,2}(t)=1-\mathcal{Z}_{1,2}(0,0,t)$ and  $S_{2,2}(t)=1-\mathcal{Z}_{2,2}(0,0,t)$. Hence from Eqs.~\eqref{PA}--\eqref{PB} we deduce that the survival probabilities $S_{1,2}(t)$ and  $S_{2,2}(t)$ evolve according to rate equations
\begin{subequations}
\begin{align}
  \label{SA}
 \frac{d S_{1,2}}{dt} &=S_{2,2}-S_{1,2}^2\\
  \label{SB}
  \frac{d S_{2,2}}{dt} &=  -S_{2,2}^2,
\end{align}
\end{subequations}
with initial conditions $S_{1,2}(0)=S_{2,2}(0)=1$. 
Note that
\begin{equation}
S_{2,2} = \sigma(S_{1,1}) =S_{1,1} =  \frac{1}{1+t}\label{eq:S2}
\end{equation}
is the survival of the single type process, given by \eqref{CBP:St}.
The governing equation \eqref{SA} therefore becomes an initial value problem 
\begin{equation}
  \label{SA:2}
 \frac{d S_{1,2}}{dt} = \frac{1}{1+t} - S_{1,2}^2, \quad S_{1,2}(0)=1.
\end{equation}
This is a soluble Riccati equation. We first simplify the non-homogeneous term by using the new time variable $\tau=\sqrt{1+t}$, which leads to 
$$
\frac{d S_{1,2}}{d\tau} = \frac{2}{\tau} - 2\tau S_{1,2}^2.
$$
We transform this into a second-order linear differential equation by setting $S_{1,2}(\tau) = \frac{1}{2\tau}\frac{B'(\tau)}{B(\tau)}$ to get
$$
B''(\tau) -\frac{B'(\tau)}{\tau} - 4B(\tau) = 0.
$$
Finally, we let $s=2\tau$ and write $B(\tau)=\frac{s}{2}A(s)$ to obtain the standard differential equation for 
modified Bessel functions \cite{NIST:DLMF}
$$
 s^2A''(s)+sA'(s) - (1+s^2)A(s) = 0.
$$
Hence the solution is a linear combination of these functions
$$
 A(s) = c_1 I_1(s) + c_2 K_1(s),
$$
where $I_p$ and $K_p$ are the modified Bessel functions of order $p$ of the first and second kind, respectively. Making the previous substitutions backward, we arrive at the solution of \eqref{SA:2} 
\begin{equation}
\label{eq:S1}
S_{1,2} = \frac{1}{\tau}\,\frac{I_0(2\tau)-cK_0(2\tau)}{I_1(2\tau) + cK_1(2\tau)}\,,\quad
\tau=\sqrt{1+t}.
\end{equation}
The initial condition $S_{1,2}(t=0)=1$ fixes the amplitude
\begin{equation}
\label{const}
c = \frac{I_0(2)-I_1(2)}{K_0(2) + K_1(2)}.
\end{equation}

\begin{figure}
\centering
\includegraphics[scale=0.6]{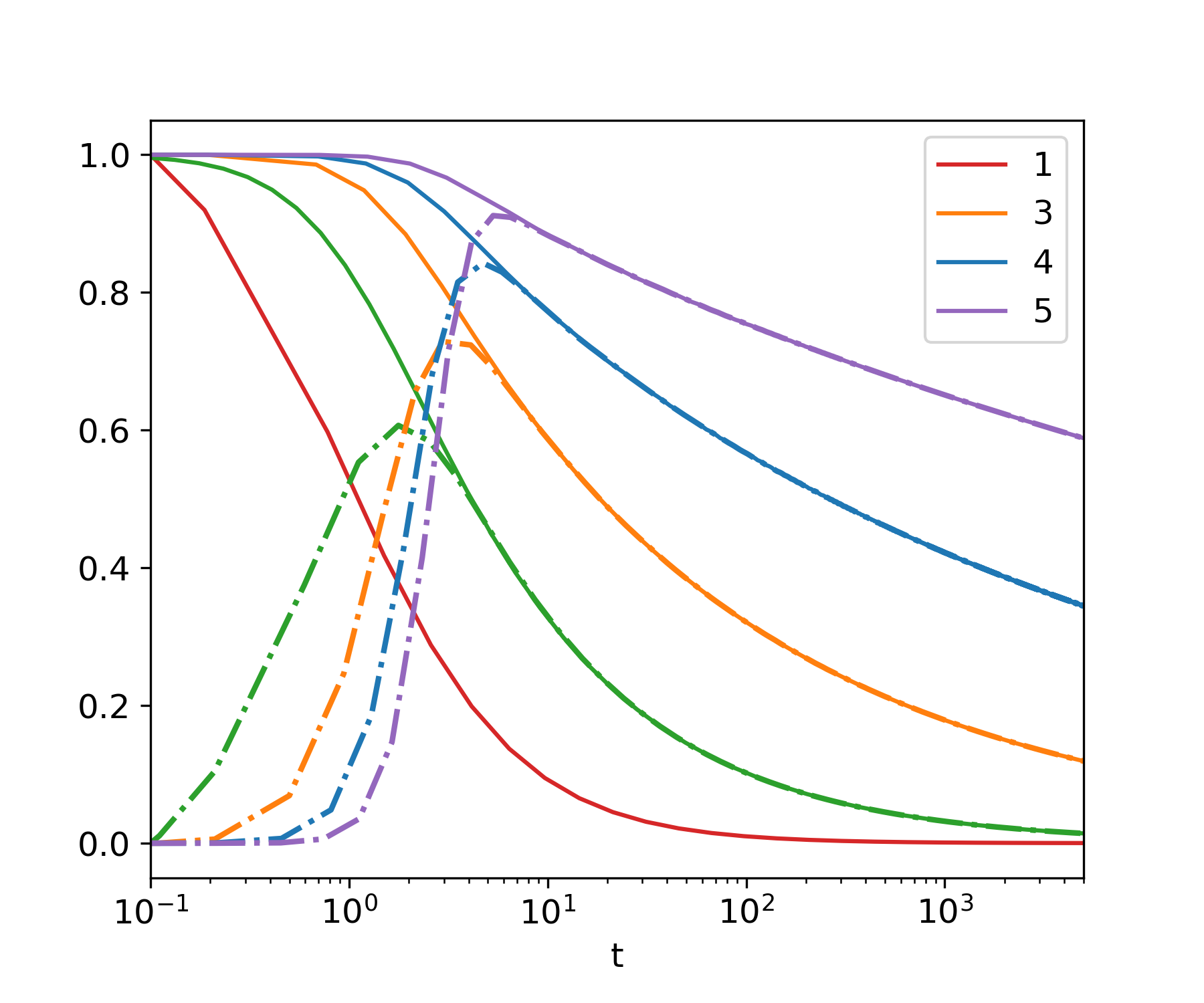}
\caption{Survival probabilities of the n-type critical process with initial condition $Z_1(0)=1$. We plot the survival probability of the entire system, $S_{1,n}(t)$ (lines) and that of just type-$n$ cells, $Q_{1,n}(t)$ (dashed-dots lines). Curves are for different types $n=1,\dots,5$. For type-1 and type-2 cells, the solutions are exact from \eqref{onetype_gen} and \eqref{PA-sol}, respectively. For other $n>2$, solutions are obtained numerically from \eqref{eq:suvi} with different initial conditions.}\label{fig:snqn}
\end{figure}

\subsection{Generating functions}

We now return to the generating functions $\mathcal{Z}_{1,2}$ and $\mathcal{Z}_{2,2}$. 
A full solution of Eqs.~\eqref{PA}--\eqref{PB} subject to the initial conditions 
\eqref{in-PA}--\eqref{in-PB} is obtained in similar way as the above solution of 
Eqs.~\eqref{PA}--\eqref{PB}. First, notice that the solution of \eqref{PB} is just the generating function \eqref{onetype_gen} of a single type but with the indexes increased by one
\begin{equation}
\label{PB-sol}
\mathcal{Z}_{2,2} 
=\sigma(\mathcal{Z}_{1,1})
= \frac{t(1-x_2)+x_2}{t(1-x_2)+1}.
\end{equation}
Plugging this into \eqref{PA} we recast it into a Riccati equation 
\begin{equation}
  \label{eq:dz1}
 \frac{d (1-\mathcal{Z}_{1,2})}{dt} = \frac{1}{(1-x_2)^{-1}+t} -(1-\mathcal{Z}_{1,2})^2.
\end{equation}
Comparing \eqref{eq:dz1} and \eqref{SA:2} we see that $1-\mathcal{Z}_{1,2}$ satisfies the same equation as $S_{1,2}$, the only distinction is in the shift of the time variable in the first term on the right-hand sides. Thus we use the variable
\begin{equation}
\label{eq:tauz}
\tau = \sqrt{t+(1-x_2)^{-1}}.
\end{equation}
and then the solution becomes 
\begin{equation}
\label{PA-sol}
\mathcal{Z}_{1,2} = 1-\frac{1}{\tau}\,
\frac{I_0(2\tau)-cK_0(2\tau)}{I_1(2\tau)+cK_1(2\tau)}\,.
\end{equation}
The initial condition $ \mathcal{Z}_{1,2}(x_1,x_2, 0) = x_1$ fixes the amplitude
\begin{equation}
\label{eq:cz}
c = \frac{I_0(2\tau_0)-(1-x_1)\tau_0 I_1(2\tau_0)}
{K_0(2\tau_0)+(1-x_1)\tau_0 K_1(2\tau_0)}\,, \quad \tau_0=(1-x_2)^{-1/2}. 
\end{equation}
For $x_1=x_2=0$ we recover the survival probabilities \eqref{eq:S1}.

Having a full solution \eqref{PA-sol}--\eqref{eq:cz} for the generating function, we can extract the probability of finding $a$ type-1 cells and $b$ type-2 cells, $P_{a,b}(t)$,  by using Cauchy's integral formula:
\begin{equation}
\label{Cauchy}
P_{a,b}(t) = \frac{1}{(2\pi i)^2}
\oint \frac{dx_1}{x_1^{a+1}}\oint \frac{dx_2}{x_2^{b+1}}\,  \mathcal{Z}_{1,2}(x_1,x_2,t).
\end{equation}
Unfortunately, explicit solutions for $P_{a,b}(t)$ are not available for arbitrary $a$ and $b$, although one can reduce the number of integrations in Eq.~\eqref{Cauchy} to one, and use numerical methods to obtain values for arbitrary $a$ and $b$, see Appendix~\ref{app:if} and \ref{app:num}. Fortunately, in many applications, a less detailed description suffices. For instance, one may be interested in the marginal distribution of type-2 cells, 
$$
P^{(2)}_{s}(t) = \mathbb{P}(Z_2(t)=s)
$$ 
or the total population size of the two-type process 
$$
\Pi^{(2)}_s(t)=\mathbb{P}(Z_1(t)+Z_2(t)=s).
$$ 
These are encoded in the generating functions
\begin{align}
  \mathcal{Z}_{1,2}(1,x,t)&=\mathbb{E}(x^{Z_2(t)}|Z_j(0)=\delta_{1j}) = \sum_{s\geq 0} P^{(2)}_s(t)x^s \label{eq:p}\\
\mathcal{Z}_{1,2}(x,x,t)&=\mathbb{E}(x^{Z_1(t)+Z_2(t)}|Z_j(0)=\delta_{1j}) = \sum_{s\geq 0} \Pi^{(2)}_s(t)x^s .\label{eq:pi-def}
\end{align}
In the next section, we derive the asymptotic behaviour of both distributions.
\section{$n$ types}
\label{multi}
In this section, we study the $n-$type critical birth-death process. This  models a population of cells that mutate at the limit mutation rate until they reach a maximal number of mutations $n$. This is
represented by the scheme 
\begin{equation*}
\begin{split}
\xymatrix{(1)+(1)           & (2)+(2)          &                    & (n)+(n)        &\\
                (1) \ar[u] \ar[r] & (2)\ar[u] \ar[r] & \cdots \ar[r]  & (n) \ar[u] \ar[r] &\emptyset}
\end{split}         
\end{equation*}
This considers a decomposable critical process in which each type can only be produced by the preceding type, and all types divide or mutate at rate one, except the maximal type-$n$ cells, which cannot mutate but die at rate one. The results are easily generalized to consider more general cases including death and multiple mutation rates- the derivations can be found in the Appendix~\ref{app:ndeath}. 

The na\"ive approach to the problem is to study $\mathbb{E}Z_i(t)$. The forward equations are $\frac{d}{dt} \mathbb{E}Z_1(t)=0$ and 
$$ \frac{d}{dt} \mathbb{E}Z_i(t)= \mathbb{E}Z_{i-1}(t)
$$
for $i=2,\dots n$, with initial condition $\mathbb{E}Z_i(0)=\delta_{i1}$. The solution is
\begin{equation}
\label{mean gen}
 \mathbb{E}Z_i(t) = \frac{t^{i-1}}{(i-1)!}.
\end{equation}
For the infinite type case ($n\to\infty$) we recover the Yule process of Section~\ref{inf-type} for the total number of cells $\mathbb{E}Z(t) = \sum_{i\ge 1} \mathbb{E}Z_i(t) = e^t$. For a finite number of types, $Z(t)=\sum_{i=1}^n Z_i(t)$, and the mean total number of cells grows exponentially $\mathbb{E}Z(t)\approx e^t$ for small $t$ but algebraically as $t^{n-1}$ when $t\to\infty$. Thus, even though the process is critical and dies out at a finite time with probability one, $\mathbb{E}Z(t)$ keeps growing forever, and therefore the na\"ive approach fails to capture one important aspect of the system. Hence instead of working with the mean number of cells, we derive asymptotic solutions for large time limits following a similar strategy introduced for the single type process.

The generating function for the $n$-type cell process starting with a single initial type-$i$ cell is
\begin{equation}
\label{gengen}
 \mathcal{Z}_{i,n}(x_1,x_2,\dots x_n,t) = \mathbb{E} \left(\prod_{j=1}^n x_j^{Z_j(t)} | Z_k(t) = \delta_{ik}\right)
\end{equation}
which can be obtained by solving the backward Kolmogorov equations
\begin{align}
  \label{eq:BK}
 \partial_t \mathcal{Z}_{i,n }=\begin{cases} \mathcal{Z}_{i,n}^2+\mathcal{Z}_{i+1,n}-2\mathcal{Z}_{i,n} & \text{for}\quad {i=1,\dots,n-1}\\ \mathcal{Z}_{i,n}^2 + 1 - 2\mathcal{Z}_{i,n} &\text{for}\quad {i=n}. \end{cases}
\end{align}
Thus, in order to obtain the interesting generating function $\mathcal{Z}_{1,n}$ for the $n$-type process starting from a single type-$1$ cell, we need to first solve the $n-1$ equations for $\mathcal{Z}_{n,n}, \mathcal{Z}_{n-1,n}, \dots, \mathcal{Z}_{2,n}$. However, since the system is decomposable, we have that $\mathcal{Z}_{i+1,j+1}=\sigma(\mathcal{Z}_{i,j})$, where $\sigma()$ is the index increase operator \eqref{eq:sigmaop}. Thus, if we know the generating function for type-$k$ cells, solving the system for $k+1$ involves solving a single additional equation. For example, for the 3-type process, $\mathcal{Z}_{3,3}=\sigma(\mathcal{Z}_{2,2})$ and $\mathcal{Z}_{2,3}=\sigma(\mathcal{Z}_{1,2})$, which we solved in the $2$-type process, and thus we only need to solve one more equation in order to obtain $\mathcal{Z}_{1,3}$. The same applies to the survival probabilities, which we consider in the following section.

\subsection{Survival probabilities}
\label{sec:n-survival}

As seen in Section~\ref{two}, it is simpler to solve the system for the survival probabilities 
\begin{equation}
     S_{i,n}(t)
 = \mathbb{P}(Z_1(t)+\dots+Z_n(t)>0| Z_k(0)=\delta_{i,k})
 = 1-\mathcal{Z}_{i,n}(0,0,\dots,0,t).
 \label{eq:sndef}
\end{equation}
Substituting into \eqref{eq:BK}, we deduce that 
\begin{align}\label{eq:suvi}
 \frac{d S_{i,n}}{dt} =\begin{cases}S_{i+1,n}-S_{i,n}^2 \quad&\text{for}\quad{i=1,\dots,n-1},\\
   -S_{n,n}^2 \quad&\text{for}\quad{i=n},\end{cases}
\end{align}
with initial conditions $S_{i,n}(0)=1$ for all $i=1,\dots,n$. For any $n$, the solutions starting with a single type-$n$ cell are given by the solutions of the single-type system up to shift in indices,
$$S_{n,n}=\sigma(S_{n-1,n-1})=\dots = \sigma(S_{2,2})=\sigma(S_{1,1})=1/(1+t).$$
To get $S_{n-1,n}$, we need to solve
\begin{equation}
\label{SA:n-1}
\frac{d S_{n-1,n}}{dt} =\frac{1}{1+t}-{S^2_{n-1,n}}.
\end{equation}
We have solved this case in Section \ref{sec:2sp} exactly, where we specified $n=2$, but the solution's form is the same for all $n$. Even though analytic solutions are invaluable, the most interesting long-time asymptotic behavior can be extracted directly from the above equation using standard methods \cite{bender2013advanced}, circumventing the exact solutions. Assume that asymptotically $S_{n-1,n}\sim (1+t)^{-\alpha}$, with $\alpha>0$, where $f(x)\sim g(x)$ means $f(x)/g(x)=1$ for $x\to\infty$. Substituting into \eqref{SA:n-1} we get 
$$
-\alpha (1+t)^{-\alpha-1} + \dots
= (1+t)^{-1} - (1+t)^{-2\alpha} + \dots
$$
where the dots refer to smaller order terms. One needs to match the coefficients of the leading order terms on the two sides, 
which leads to a contradiction for both $2\alpha>1$ and $2\alpha<1$, thus $2\alpha=1$. That is, in the leading order, the right-hand side of  \eqref{SA:n-1} should vanish faster than $1/(1+t)$.  This gives
\begin{equation}
\label{s_leading}
S_{n-1,n} \sim (1+t)^{-1/2}.
\end{equation}
To get to higher orders we just add new terms to \eqref{s_leading} of the form $b{(1+t)}^{-\beta}$ one by one with smaller values of $\beta$ each time, substitute into \eqref{SA:n-1} and match the coefficients. For the first 3 terms in the $t\to\infty$ asymptotic limit we get
\begin{equation}
  \label{SA:2-two+}
S_{n-1,n} \sim (1+t)^{-1/2}+\tfrac{1}{4}(1+t)^{-1}+\tfrac{3}{32}(1+t)^{-3/2} .
\end{equation}
Here $f(x)\sim g(x)+h(x)$ means $[f(x)-g(x)]/h(x)=1$ for $x\to\infty$.
Note that we performed the expansion in powers of $1+t$ for convenience, but one could simply expand the result in powers of $t$ instead. This expansion can be also derived from the exact formula \eqref{eq:S1} using large argument asymptotic expansion of the Bessel functions (Appendix~\ref{app:formulas}).

\begin{figure}
    \centering
    \includegraphics[scale=0.6]{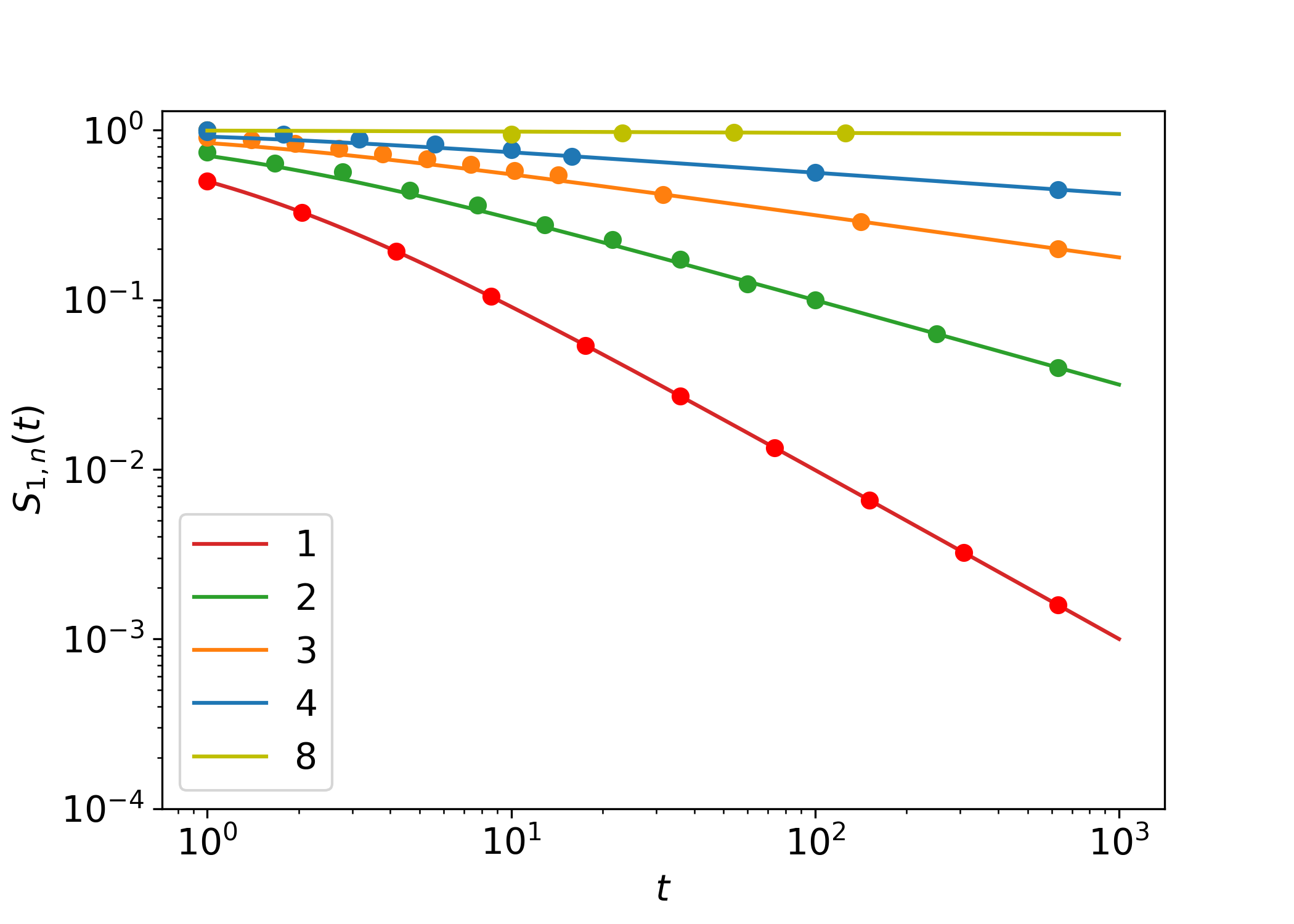}
    \caption{ Survival probability of the n-type process, for $n=1,2,3,4,8$. Second-order asymptotic solutions obtained from \eqref{eq:sn-as} (lines) are plotted together with results from simulations (dots).}
    \label{fig:extinction N type}
\end{figure}

Continuing to the next type, we have the differential equation for $S_{n-2,n}$ in \eqref{eq:suvi} where $S_{n-1,n}$ is replaced by its above expansion. This procedure gives the leading order for all types by setting the time derivative to zero. In the leading order, we get that $S_{n-j,n}\sim (1+t)^{-\chi_{j+1}}$ with $\chi_j=2^{1-j}$. With a little more work, one obtains the next correction
\begin{equation*}
    S_{n-j,n}\sim
{(1+t)}^{-\chi_{j+1}}+ 2^{-j-1}{(1+t)}^{-\tfrac{1}{2}-\chi_{j+1}}
\end{equation*}
for $j=1,\dots,n-1$. The most interesting survival probability is $S_{1,n}$, which gives the probability that the $n$-type system starting with a single type-1 cell is not empty at time $t$. Asymptotically, this is given by
\begin{equation} 
 \label{eq:sn-as}
 S_{1,n}\sim (1+t)^{-\chi_n}+\frac{\chi_n}{2}{(1+t)}^{-\tfrac{1}{2}-\chi_n}
\end{equation}
for $n\ge 2$, and for $n=1$ the first term is exact.
One can consider higher-order terms by successively adding terms and matching
coefficients. 
Figure \ref{fig:extinction N type} shows that the second-order asymptotic accurately describes the behaviour.

\subsection{Generating functions}
\label{sec:n-gener}

We now attempt to find solutions for the generating functions for the number of cells. 
We rewrite the Kolmogorov equations \eqref{gengen}
into 
\begin{align}\label{eq:numi}
 \frac{d (1 - \mathcal{Z}_{i,n})}{dt} =\begin{cases}
 (1 - \mathcal{Z}_{i+1,n})-(1 - \mathcal{Z}_{i,n})^2 \quad&\text{for}\quad{i=1,\dots,n-1},\\
   -(1 - \mathcal{Z}_{i,n})^2 \quad&\text{for}\quad{i=n},\end{cases}
\end{align}
which for $1-\mathcal{Z}_{{j,n}}$ are identical to equations for the survival probability \eqref{eq:suvi},
but with initial conditions $1-\mathcal{Z}_{j,n}(x_1,\dots,x_n,0) = 1-x_j$.
This is not surprising since the survival probability is $S_{i,n}(t) = 1 - \mathcal{Z}_{i,n}(0,\dots0,t)$. 

We again treat this system by first considering the generating function for the system starting with a single type-$n$ cell, which corresponds to the single type case. Hence
$$
 1 - \mathcal{Z}_{n,n} = \frac{1}{t+ (1-x_n)^{-1}}, 
$$
which is the same as if we replace $1+t$ in $S_{n,n}$ by $t+ (1-x_n)^{-1}$. Starting with a single type-$(n-1)$ cell we have
\begin{equation}
\label{dist:n-1}
 \frac{d (1 - \mathcal{Z}_{n-1,n})}{dt} =
 \frac{1}{t+ (1-x_n)^{-1}}-(1 - \mathcal{Z}_{n-1,n})^2.
\end{equation}
To get a nontrivial large-time behaviour we need $(1-x_n)^{-1}\propto t$. Noting the similarity between  \eqref{dist:n-1} and \eqref{SA:n-1}, we assume a power behavior for $1-\mathcal{Z}_{n-1,n}$ which fixes the exponent and leads to
$$
1-\mathcal{Z}_{n-1,n} \sim {\left({t+ (1-x_n)^{-1}}\right)}^{-1/2}.
$$ The same result can be also obtained directly from the explicit solution \eqref{PA-sol}  for the generating function of type-2 cells  $\mathcal{Z}_{1,2}(1,x_2,t)$ by taking the large argument asymptotic of the modified Bessel functions \cite{Antal:2011}. 

Continuing this procedure we get that, to the leading order, 
$$
1-\mathcal{Z}_{n-2,n} \sim {\left({t+ (1-x_n)^{-1}}\right)}^{-1/4},
$$
and so on. In particular, the generating function starting with a single type-1 cell is given by
\begin{equation}
\label{eq:aszn}
1-\mathcal{Z}_{1,n} \sim  {\left({t+ (1-x_n)^{-1}}\right)}^{-\chi_n} .
\end{equation}
Thus we see that the same exponents, $\chi_n=2^{1-n}$, that describe the survival probabilities give the asymptotic behaviour of the generating functions. Note that in the leading order, the generating functions only depend on the last type of cells $x_n$. This is not surprising if we note that the survival probability of type-$k$ cells is the square root of the survival probability of type-$k-1$ cells. Thus the system becomes dominated by the last type. That is, the generating function for the last type is asymptotically the same as the generating function for the total number of cells,
\begin{equation}
    \mathcal{Z}_{1,n}(1,\dots,1,x_n,t)\sim \mathcal{Z}_{1,n}(x_n,\dots,x_n,t) .
\label{eq:asgn}
\end{equation}
If we take \eqref{eq:asgn} at $x_n=0$ we see that the survival probability of just the type-$n$ cells
\begin{equation}
Q_{i,n}(t) = 
 \mathbb{P}(Z_n(t)>0 | Z_k(0)=\delta_{i,k}) 
 = 1-\mathcal{Z}_{i,n}(1,\dots,1,0,t)
\label{eq:qn}
\end{equation}
is asymptotically the same as the survival probability of $S_{1,n}(t)$ of the whole system, that is
\begin{equation}
Q_{1,n}(t) \sim S_{1,n}(t).
\label{eq:qn_sim_sn}
\end{equation}
This can also be seen in Figure \ref{fig:snqn}, where we plotted exact ($n=1,2$) and numerical ($n>2$) solutions for $S_{1,n}(t)$ and $Q_{1,n}(t)$.
It is clear from \eqref{eq:BK} that $Q_{i,n}(t)$
is governed by the same equations as the survival probability $S_{i,n}(t)$ of the entire system \eqref{eq:suvi}
but with initial conditions $Q_{i,n}(0)=0$ for $i=2,\dots,n-1$ and $Q_{n,n}(0)=1$. In Figure \ref{fig:snqn} we see that subsequent types arise fast but disappear slowly, as the system gets dominated by the last type.

\subsection{Cell number distributions}

We now derive the asymptotic distribution of the number of type-$n$  cells 
\begin{equation*}
P^{(n)}_s(t)= \mathbb{P}(Z_n(t)=s|Z_k(0)=\delta_{k,1})
\end{equation*}
which is encoded in the generating function
\begin{equation}\label{eq:z1n}
\mathcal{Z}_{1,n}(1,\dots,1,x,t) 
= \mathbb{E} (x^{Z_n}|Z_k(0)=\delta_{k,1})
= \sum_{s\geq0}P^{(n)}_s(t) x^s .
\end{equation}
To get non-trivial large-time behaviour, we need to condition the generating function on the survival of type-$n$ cells.
Following the procedure of Section~\ref{sec:one-type}, we express the conditional generating function as
\begin{equation}
\label{twotypelim}
 \mathbb{E}(x_n^{Z_n(t)}|Z_n(t)>0) 
 = \frac{\mathcal{Z}_{1,n}(1,\dots,1,x_n,t)-P^{(n)}_0(t)}{1-P^{(n)}_0(t)}
\end{equation}

As in the single type case, we obtain a nontrivial scaling by using the scaling variables $x_n=1-p/t$ and $y=s/t$ when taking $t\to \infty$ with $p$ and $y$ constants. Substituting the leading order asymptotic expressions for the survival probability of type-$n$ cells from \eqref{eq:qn_sim_sn} and \eqref{eq:sn-as}
\[
1 - P^{(n)}_0(t) = Q_{1,n}(t) \sim S_{1,n}(t) \sim t^{-\chi_n}
\]
and that of the generating function from \eqref{eq:aszn}
$$
 1-\mathcal{Z}_{1,n}(1,1-p/t,t)
 \sim (t+t/p)^{-\chi_n}.
$$
we get that 
\[ \mathbb{E}(x_n^{Z_n(t)}|Z_n(t)>0) \to 1 - {\left(\frac{p}{p+1}\right)}^{\chi_n}.\]
To obtain this convergence in a different way requires the following convergence to the density of a random variable $Y_n$ 
$$
 t\frac{P^{(n)}_s(t)}{Q_{1,n}(t)} \sim t^{1+\chi_n} P^{(n)}_{yt}(t)\to f_{Y_n}(y)
$$
so the generating function becomes a Riemann integral in the $t\to\infty$ limit:
\begin{equation*}
\begin{split}
\mathbb{E}(x_n^{Z_n(t)}|Z_n(t)>0)
&= \sum_{s\geq 1} \frac{P^{(n)}_s(t)}{Q_{1,n}(t)} x^s = 
\sum_{s\geq 1} t\frac{P^{(n)}_s(t)}{Q_{1,n}(t)} {(1-p/t)}^{yt} \frac{1}{t}\\
&\to \int_0^\infty f_{Y_n}(y) e^{-py} dy 
= \mathbb{E} e^{-pY_n}
= 1-{\left(\frac{p}{p+1}\right)}^{\chi_n}.
\end{split}
\end{equation*}
Hence we obtained a convergence in distribution
$$
 \frac{Z_n(t)}{t}|\{Z_n(t)>0\}
 \to Y_n.
$$

Inverting the above Laplace transform $\mathbb{E} e^{-pY_n}$ we express the density of the limit variable via the confluent hypergeometric function \cite{NIST:DLMF}
\begin{equation}
\label{Pzn}
f_{Y_n}(y)
=\chi_n F(1 + \chi_n; 2; -y).
\end{equation}
Therefore, the scaling form of the type-$n$ distribution is
\begin{equation}
\label{Pst-scaling_og}
P^{(n)}_{s}(t)\approx \chi_n t^{-1-\chi_n}F(1 + \chi_n; 2; -s/t).
\end{equation}
For $n=1$, since $\chi_1=1$ and $F(2;2;-x)=e^{-x}$, we recover the exponential limit of the single type case given in \eqref{onetype_scaling}.

Finally, for $n\ge 2$, by taking the large argument asymptotic of the confluent hypergeometric function (13.7.2 in \cite{NIST:DLMF}), we obtain the large $y=s/t$ tail of the distribution.
\begin{equation}
\label{tail:n}
P^{(n)}_{s}(t) \approx \frac{\chi_n}{\Gamma(1-\chi_n)}\, s^{-1-\chi_n}
\quad\text{when}\quad  t\ll s\ll t^{\tfrac{n-1}{1-\chi_n}}.
\end{equation}
The algebraic decay for $n\ge 2$ is in stark contrast with the exponential decay of the first type cells, $n=1$.
Surprisingly, the tail is not only algebraic but also stationary. The algebraic tail is describes the behaviour for a limited range of $s$ values. The lower bound $s\ll t$ comes from the large $y=s/t$ expansion. The upper bound is more subtle, and comes from noticing that the algebraic tail would imply an infinite mean, in contradiction with the finite mean we obtained in \eqref{mean gen}. To reconcile this, we set an upper bound, $s_*$, and estimate its order of magnitude,
$$
 \mathbb{E} Z_n \propto t^{n-1} \propto \int_1^{s_*} s\, s^{-1-\chi_n} ds
 \propto s_*^{1-\chi_n}
$$
which gives $s_* \propto t^{\tfrac{n-1}{1-\chi_n}}$ as we announced.

\begin{figure}
\centering
\includegraphics[scale=0.6]{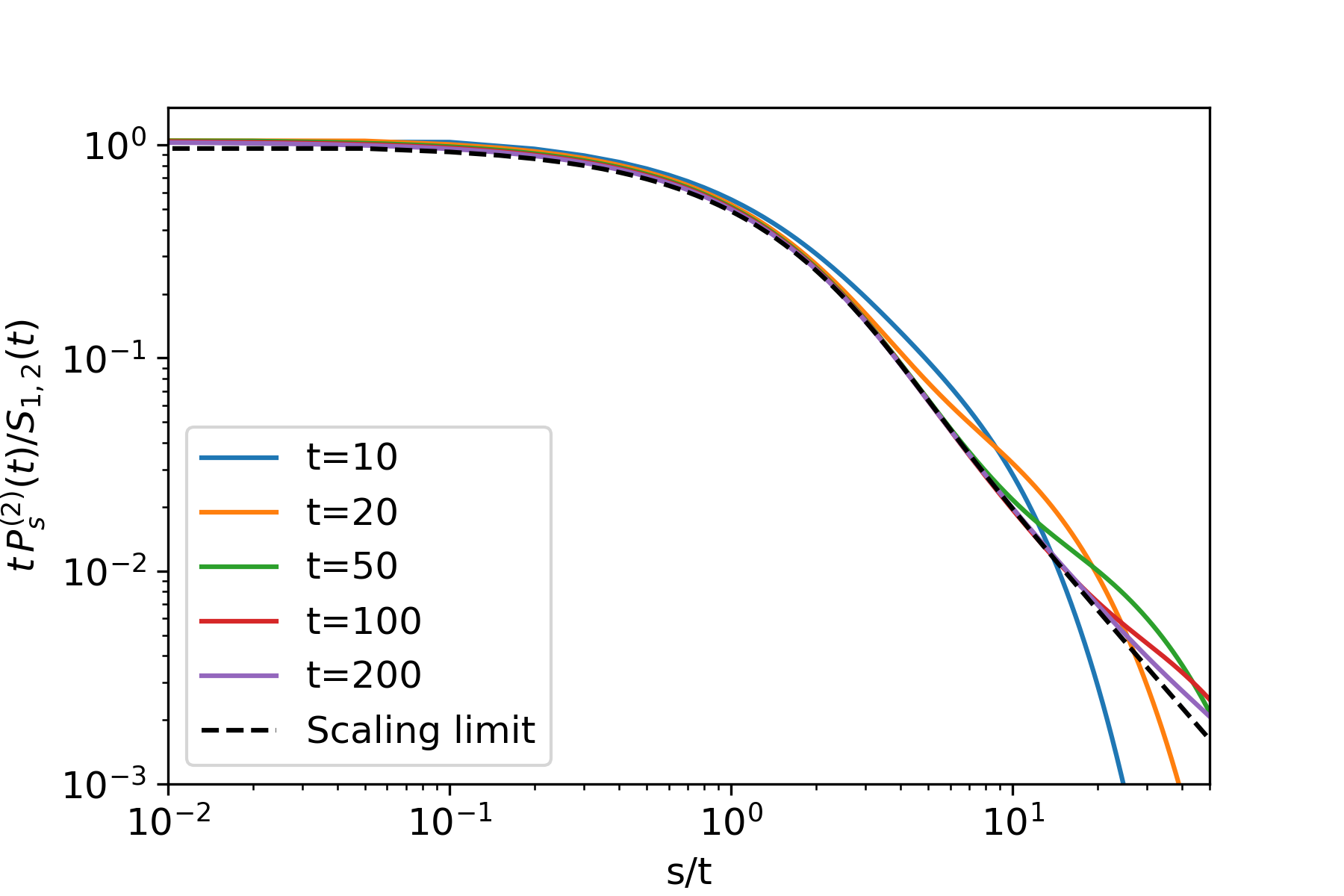}
\caption{Scaling of the probability $P_s^{(2)}(t)$ of finding $s$ type-2 cells at time $t$; different lines indicate different times $t$. Probabilities are calculated numerically
from the exact generating function \eqref{PA-sol} via the Inverse Fast Fourier Transform algorithm, see Appendix \ref{app:num}. In the double limit $t,\ s \to \infty$, with $s/t$ constant, the scaled distributions converge to the scaling limit given by \eqref{Pst-scaling_og}, depicted by dashed line.} 
\label{fig:num}
\end{figure}

\begin{figure}
\centering
\includegraphics[scale=0.6]{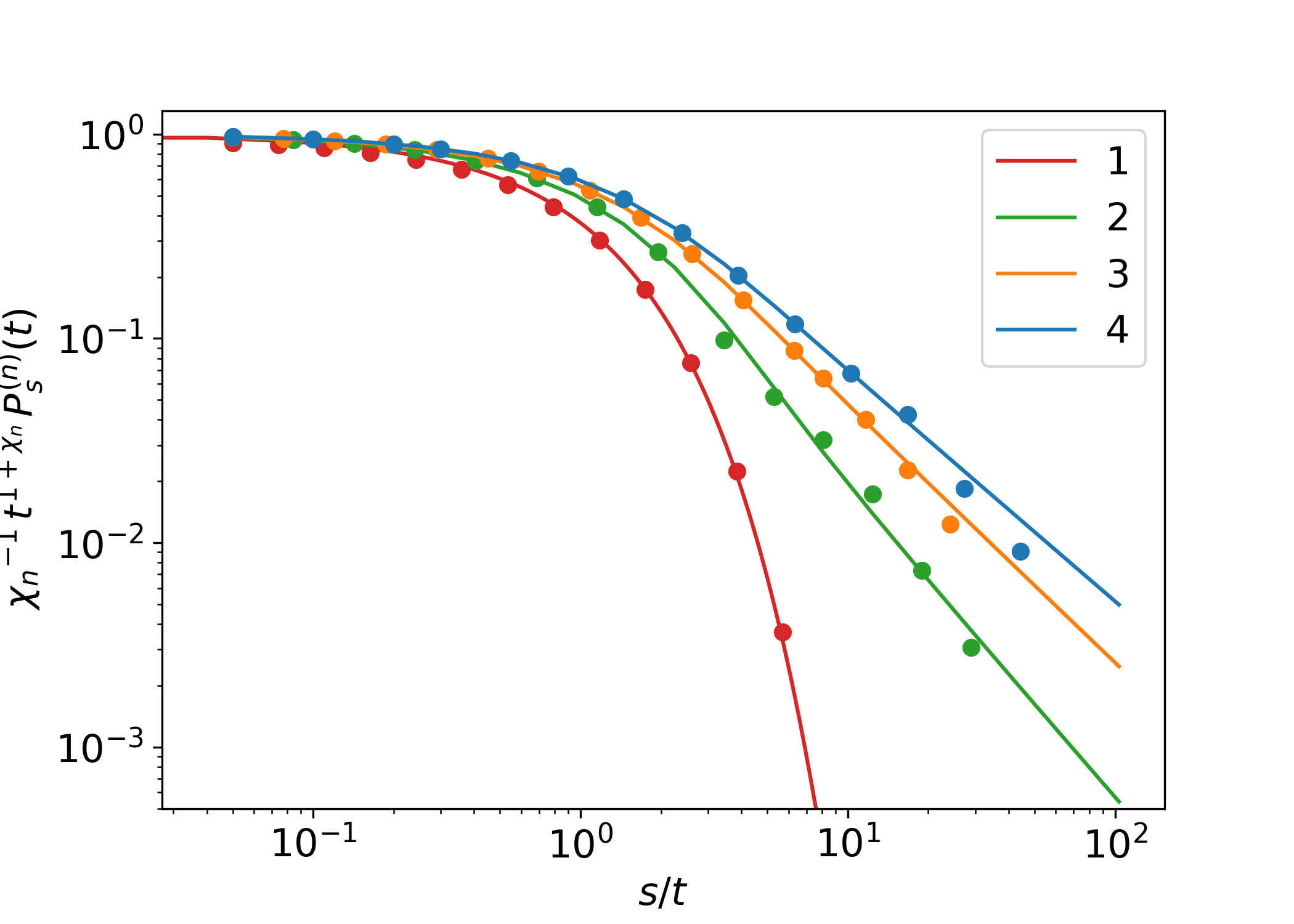}
\caption{The scaled number distributions $\chi_n^{-1} t^{1+\chi_n} P^{(n)}_s(t)$ from simulations (dots) are compared to the theoretical prediction $F(1+\chi_n; 2; -s/t)$, cf. Eq.~\eqref{Pst-scaling_og}, as a function of $s/t$, where we fix $t=20$. Curves are for $n=1,2,3,4$. Note the power law decay for $n\ge 2$ as opposed to the exponential decay for the single type case ($n=1$).}
\label{fig:scaling}
\end{figure}

The validity of the scaling limit \eqref{Pst-scaling_og} is illustrated in Figure \ref{fig:num} for $n=2$ via comparison to numerical solutions. One can see how the range of the algebraic tail expands with time. In Figure \ref{fig:scaling}, the scaled number distribution $\chi_n^{-1} t^{1+\chi_n} P^{(n)}_s(t)  =  F\big(1+\chi_n; 2; - s/t\big)$ as given by \eqref{Pst-scaling_og} is compared to simulations as a function of $s/t$, where we chose $t=20$. For $n>1$ the asymptotic solution matches simulations, but eventually overestimates the probability for large $s$. For larger $t$, the asymptotic solution is better at the large $s$ limit but under-estimates for small $s$. Within the derived range, the cell number distribution is well described by  stationary tail $(\text{size})^{-1-\chi_n}$. 

We have derived the asymptotic behaviour of the last type of cells in the $n$-type process. Since the system is decomposable, in order to get the distribution of the previous types $k=1,2,\dots,n-1$, we simply need to stop the process at the $k$th type.
In biological applications, we may also be interested in the total number of cells $Z=Z_1+\dots+Z_n$, with distribution
\begin{equation*}
\Pi^{(n)}_s(t)= \mathbb{P}(Z(t)=s| Z_k(0)=\delta_{1,k}).
\end{equation*}
This is encoded in the generating function
\begin{equation}\label{eq:pin}
\mathcal{Z}_{1,n}(x,\dots,x,x,t) = \sum_{s\geq0}\Pi^{(n)}_s(t) x^s ,
\end{equation}
hence we see by \eqref{eq:asgn} that it is asymptotically the same as the distribution of the last type of cells, 
$$\Pi^{(n)}_s(t)\sim P^{(n)}_s(t).$$
Thus, the above analysis gives us access to both the asymptotic behaviour of the total number of cells, as well as the behaviour of each individual cell type.

\subsection{Arrival and exit times}\label{sec:arrn}

Let us study when new cell types appear and disappear from the system. It is easier to work with the infinite type version of the model for this question, otherwise, we need to assume that the type in question is not greater than $n$. For convenience, we study when type-$n$ cells appear or disappear, but the results are of course valid for any type, not only the last.

For the pure birth-mutation process, the exit time of types is straightforward.
Let
$$
 E_n = \sup\{ t\ge 0 : Z_n(t)>0 \}
$$
denote the time of extinction of type-$n$ cells.
For the birth-mutation process, notice that
$$
\{E_n>t\} = \{Z_1(t)+\dots+Z_n(t)>0\}.
$$
Hence the distribution of $E_n$ when starting from a single type-$1$ cell is given by the total survival of the $n$-type process, which we derived asymptotically in \eqref{eq:sn-as}.

Let us now turn to the arrival time 
$$
 T_n = \inf\{ t\ge 0 : Z_n(t)>0\}
$$
when the first type-$n$ cell appears. 
We are interested in its distribution starting with a single type-$i$ cell
$$
h_{i,n}(t) = \mathbb{P}(T_n>t|Z_k(0)=\delta_{ik}).
$$ 
To derive an equation for this quantity consider a modified system where type-$n$ cells neither divide nor die, just stay alive forever. Hence their generating function, when starting with a single type-$n$ cell, stays constant ${\mathcal{Z}}_{n,n}=x_n$, but all other equations for $i<n$ remain the same in \eqref{eq:BK}. 
The existence of a type-$n$ cell in the modified system then indicates that they were produced also in the original system. Hence $h_{i,n}(t)={\mathcal{Z}}_{i,n}(1,\dots,1,0,t)$, and thus
\begin{equation}
    \frac{dh_{i,n}}{dt}  = h_{i,n}^2 - 2h_{i,n} + h_{i+1, n} \label{eq:arrexback}
\end{equation} 
with initial condition $h_{i,n}(0)=1$ for $i=1\dots,n-1$ and $h_{n,n}(t)\equiv 0$. In terms of $g_{i,n}(t)=1-h_{i,n}(t)=\mathbb{P}(T_n\le t|Z_k(t)=\delta_{ik})$, equation
\eqref{eq:arrexback} takes a simpler form
\begin{equation}
  \frac{d g_{i,n}}{dt}  =-g_{i,n}^2  +  g_{i+1, n} \label{eq:gin}
\end{equation}
with initial condition $g_{i,n}(0)=0$ for $i<n$ and $g_{n,n}(t)\equiv 1$ for all $t$.

For the arrival of the first type-$2$ cell, the above equation becomes 
\begin{equation}
\label{arrivaltime2_eq}
  \frac{d g_{1,2}}{dt}  =-g_{1,2}^2  +  1
\end{equation}
with solution
\begin{equation}
    g_{1,2}(t) = \tanh{t} \label{eq:twoarr}
\end{equation}
Hence the first type-2 cell arrives on average at
$$
 \mathbb{E}T_2 = \log 2.
$$
Note also that 
\begin{equation}
    \label{arrivaltime2}
 h_{1,2}(t) = 1-g_{1,2}(t) = \mathbb{P}(T_2>t|Z_k(0)=\delta_{1k}) = \frac{2}{1+e^{2t}}
\end{equation}
have the same form as for the analogous supercritical process \cite{nicholson2022mutation}.

Now we can use this solution for the arrival of type-3 cells, by noting that $g_{i +1,n}=\sigma( g_{i,n-1})$ and we get
\begin{equation}
     \label{eq:threearr}\frac{d g_{1,3}}{dt} = -g_{1,3}^2 + \tanh{t} .
\end{equation}
The solution of \eqref{eq:threearr} can be expressed as a lengthy combination of hypergeometric functions. 
For subsequent types, no exact solutions are available. 

Let us study instead the more general birth-death process described by scheme \eqref{fig:bd_multid}. The above method presented for the birth-mutation process stays valid, and for a constant mutation rate $\nu$ for all types, equation \eqref{eq:gin} becomes
\begin{equation}
\frac{d  g_{i,n}}{dt} = - g_{i,n}^2 + \nu g_{i+1, n}\label{eq:gin death}
\end{equation} 
with initial conditions $g_{i,n}(0)=0$ and $g_{n,n}\equiv 1$. 
The simplest non-trivial property of this process is the probability that the first type-$n$ mutant eventually arrives starting from a single type-$1$ cell,
$$
g_{1,n}(\infty):=\lim_{t\to \infty} g_{1,n}(t).
$$ 
For the simple birth-mutation process ($\nu=1$) this quantity is trivial: $g_{1,n}(\infty)=1$ for all $n$, namely all types arrive eventually with probability $1$. 
This is not the case in the more general birth-death case where, by setting the left-hand side of \eqref{eq:gin death} to zero, we obtain that 
$$
g_{1,n}(\infty)=\nu ^{1-\chi_n}.
$$
If we denote by $M$ the maximal type that ever appears, then what we found is that $\mathbb{P}(M\ge n)=g_{1,n}(\infty)=\nu ^{1-\chi_n}$. It gets easier for larger types to appear, in the sense that $\mathbb{P}(M\ge n+1|M\ge n) = \nu^{2^{-n}}\to 1$ as $n\to \infty$. This also implies that the mean number of types that ever appear is infinite, $\mathbb{E}M=\infty$.

Since $g_{2,2}\equiv 1$, we can get an explicit solution
for the arrival of the first type-2 cell
\[
g_{1,2}(t)=\sqrt{\nu}\tanh\sqrt{\nu} t.
\]
which generalises \eqref{eq:twoarr}.

\begin{figure}
\centering
\includegraphics[scale=0.35]{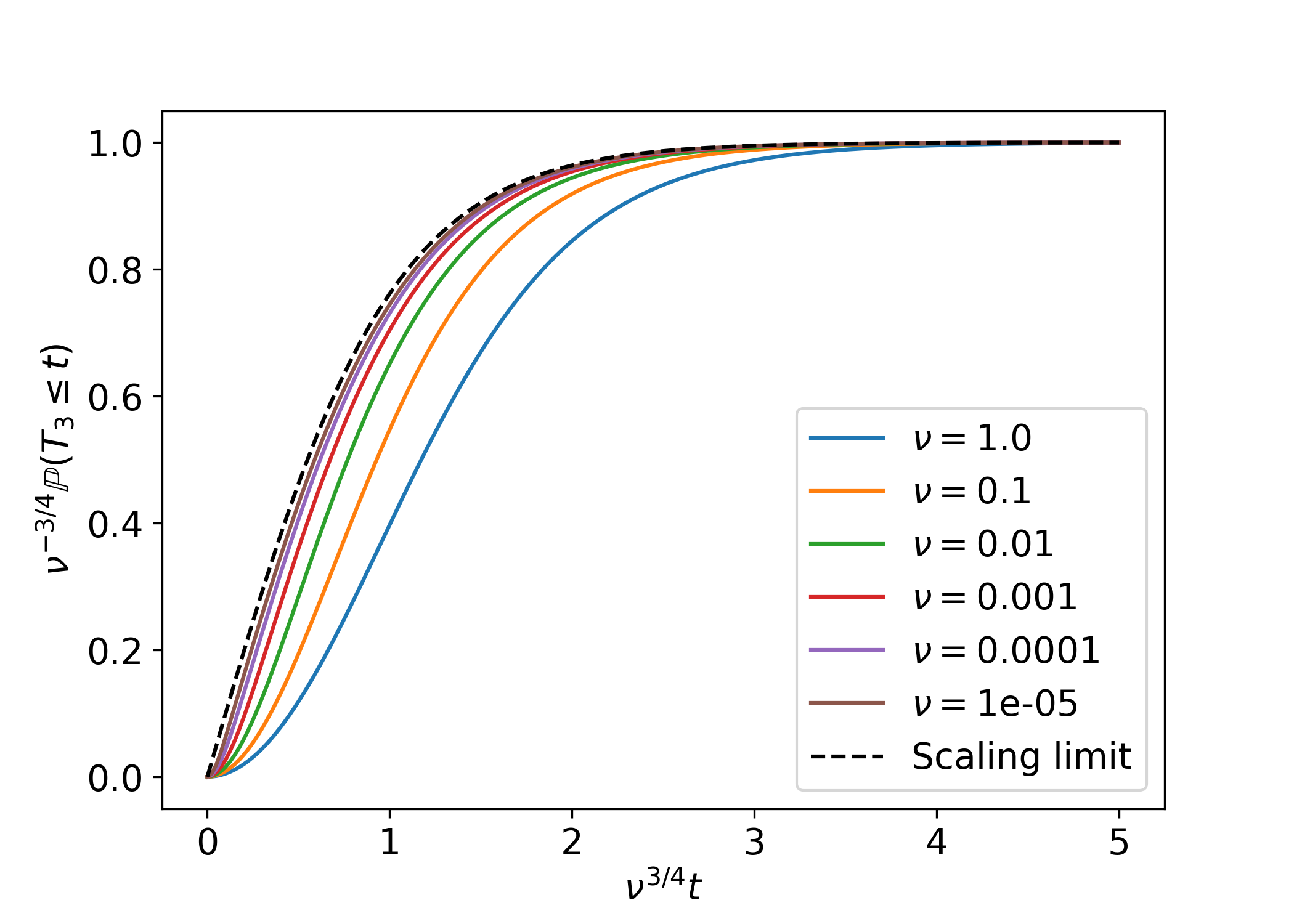} \includegraphics[scale=0.35]{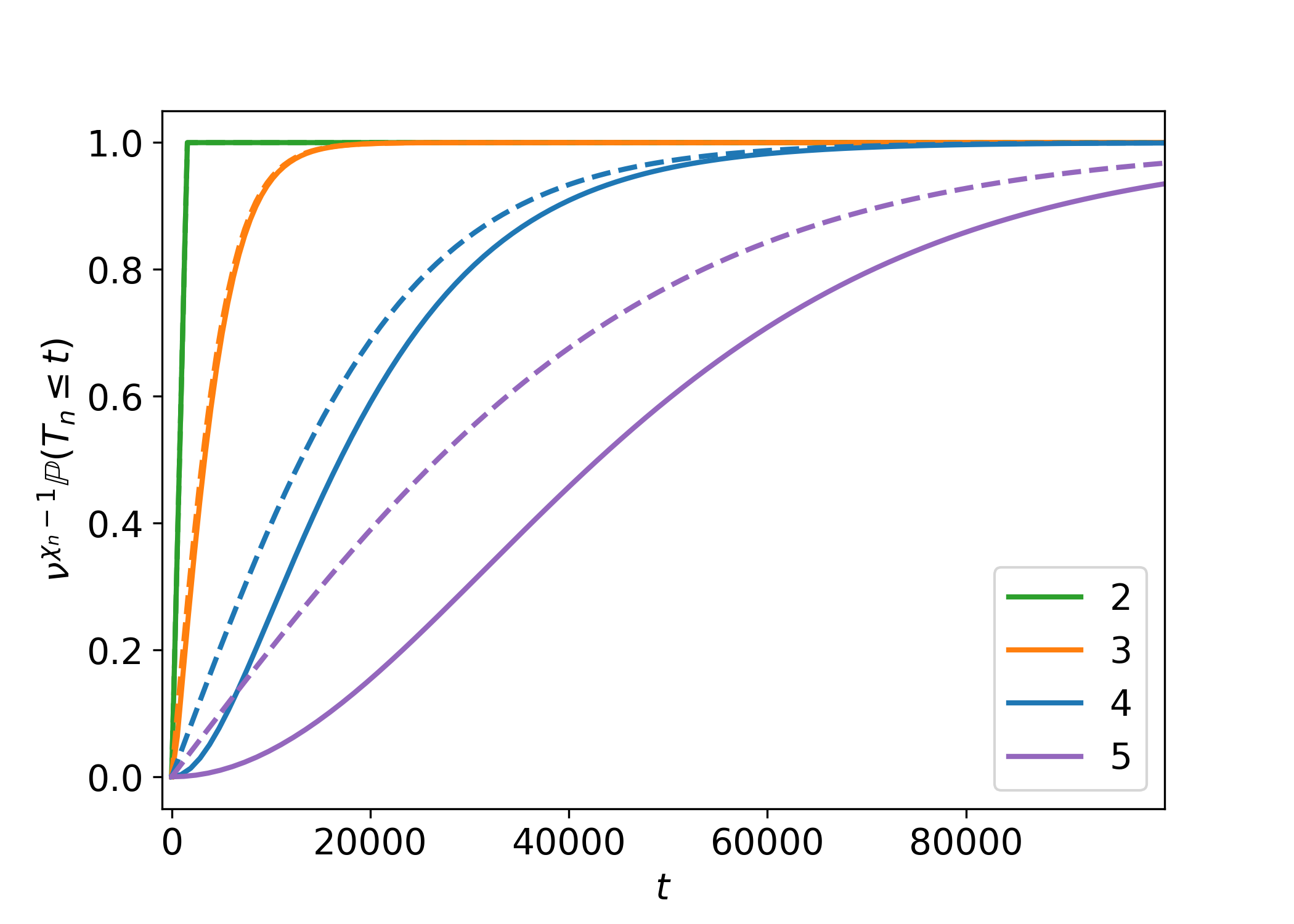} 
\caption{On the left, the scaled probability of arrival of the first type-$3$ cell, $\nu^{-3/4}\mathbb{P}(T_3\leq t)$ in terms of $\nu^{3/4}t$, obtained numerically from \eqref{eq:gin death} for different $\nu$ is compared to the scaling limit \eqref{smallnu}. On the right, the normalized arrival probability $\nu^{\chi_n-1}\mathbb{P}(T_n\leq t)$ of different types obtained numerically from \eqref{eq:gin death} are compared to the corresponding scaling limit \eqref{smallnu} (dashed line) for $\nu=10^{-5}$ and $n=2,\dots,5$.}
\label{fig:nu}
\end{figure}

For the arrival of the first type-3 cell, using that $g_{2,3}=\sigma(g_{1,2})=g_{1,2}$, we have that
\begin{equation}
\frac{d  g_{1,3}}{dt} = - g_{1,3}^2 + \nu\sqrt{\nu}\tanh\sqrt{\nu} t.
\label{saling-smallnu}
\end{equation} 
The initial condition is $g_{1,3}(0)=0$. Let us normalize $g_{1,3}$ by its limiting probability $g_{1,3}(\infty)=\nu^{3/4}$. That is, we introduce $\tilde{g}_{1,3}:=\tfrac{g_{1,3}}{g_{1,3}(\infty)}$
to get
\[
\frac{1}{g_{1,3}(\infty)}\,\frac{d \tilde{g}_{1,3} }{dt}= -  \tilde{g}_{1,3}^2 + \tanh{\sqrt{\nu }t} .
\]
Next, we re-scale time, $t\to \tilde{t}=t g_{1,3}(\infty)=t \nu^{3/4}$, and define 
$$
k_{1,3}(\tilde{t})=\tilde{g}_{1,3}\left(\tilde{t} \nu^{-3/4}\right)
= \nu^{-3/4} g_{1,3}(t)
$$
to arrive at
\[\frac{d k_{1,3}} {d \tilde{t}}= - k_{1,3}^2 + \tanh{\nu^{-1/4} \tilde t}.\]
In the limit where $\nu\to 0$ we get
\[
 \frac{d k_{1,3}} {d \tilde{t}}\approx - k_{1,3}^2 + 1 ,
\]
which is the same differential equation we have for $g_{1,2}$ for $\nu=1$ in \eqref{arrivaltime2_eq} with the same initial condition $k_{1,3}(0)=0$. Hence $k(\tilde t) = \tanh \tilde t$ and then
$$
g_{1,3}(t) \approx \nu^{3/4} \tanh{  \nu^{3/4} t}.
$$
The same procedure can be applied to obtain the arrival distribution of all types,
\begin{equation}
\label{smallnu}
g_{1,n}\left(t \right)\approx g_{1,n}(\infty) 
\tanh (g_{1,n}(\infty) t)
=\nu^{1-\chi_n}\tanh{\nu^{1-\chi_n} t}
\end{equation}
which can be then verified by induction. In Figure \ref{fig:nu} we observe that, for a fixed type, the asymptotic agrees with the behaviour in the $\nu\to0$ limit. However, for a fixed $\nu$, the approximation becomes worse as we consider more types.

\section{Examples and applications}\label{sec:exs}
We now discuss example applications of the results for studying the evolution of microbes and tumour cells at the limiting mutation rate and relate them to experimental work.
\subsection{Colony size and number of generations until EEX in microbes}

In bacteria, haploid and diploid yeast, numerous experimental projects have investigated EEX by crossing cell lines with different mutator alleles \cite{morrison1993pathway,fijalkowska1996mutants, herr2011mutator, herr2014dna, soriano2021expression}. The observable quantities of such experiments are the CanR mutation rate of the cell lines (number of mutations in CanR gene per division), and the number of viable colonies after a given number of generations. 
The main goal is to identify the maximum mutation rate. We have seen that defining the limiting mutation rate is conceptually straightforward: if the sum of death and mutation rate is at least equal to the division rate, and there is a finite number of mutations that can accumulate, then EEX will occur with probability one. However, for how long and how large the colonies can grow depends on a number of factors, including the number of mutations that can accumulate before the lethal mutation. These are questions of interest for experiment design, which we can answer using the model proposed. 

\begin{figure}
\centering
\includegraphics[scale=0.35]{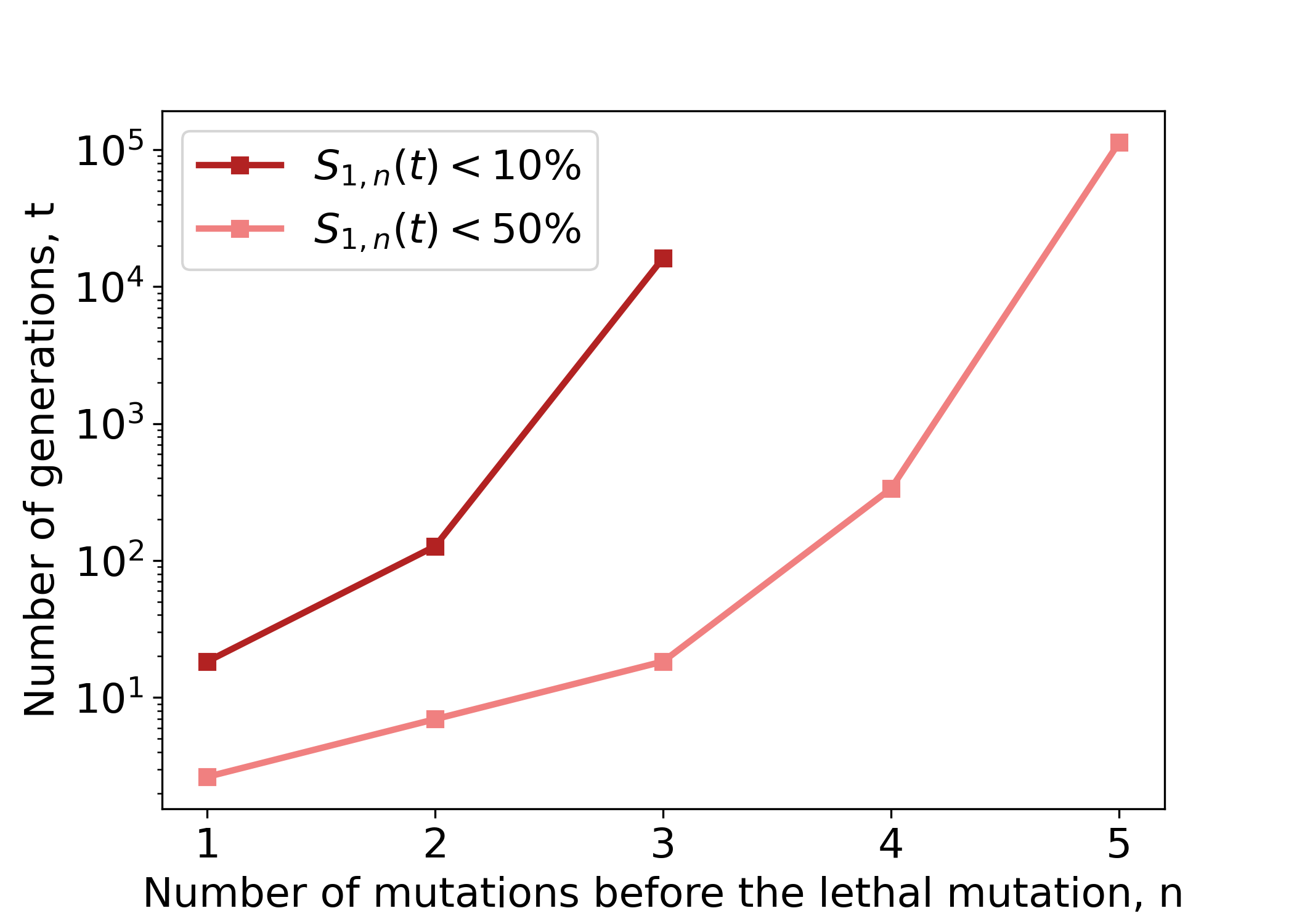} 
\includegraphics[scale=0.35]{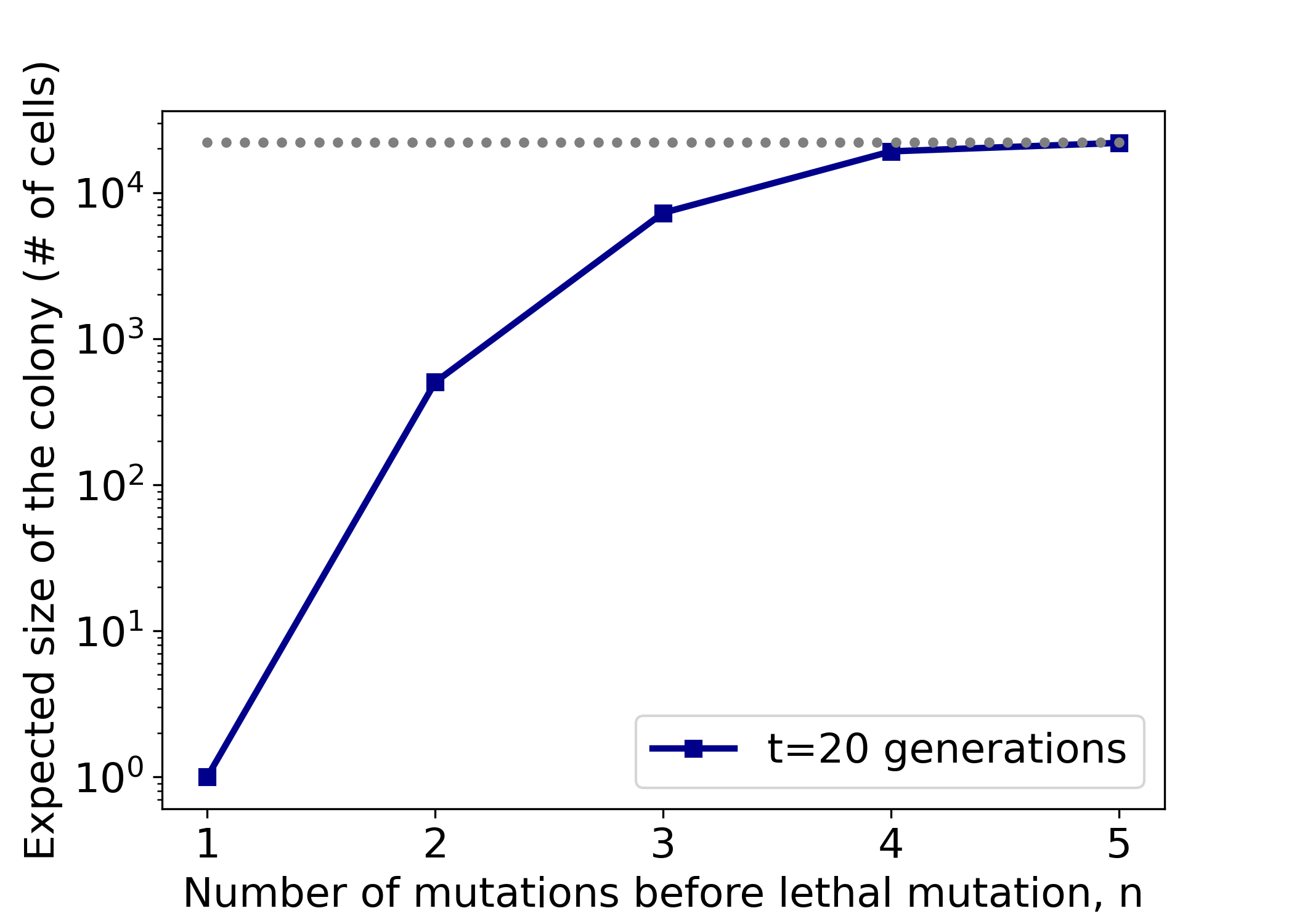} 
\caption{(A) Number of generations until the survival probability is less than $10\%$ and $50\%$,  as a function of the maximal number of mutations that can accumulate, calculated from \eqref{eq:sn-as}. (B) The expected size of colonies started with one single cell after 20 generations, as a function of the maximal number of mutations that can accumulate, calculated from \eqref{eq:pin}. The dashed grey line shows the expected size of a population growing exponentially with no maximal number of mutations.}
\label{fig:ex1}
\end{figure}

If we set the birth rate to 1 in the no-death model, we can interpret time in units of cell divisions or generations. We can immediately compute the probability that a colony initiated by a single cell will survive after $t$ generations  immediately from Equation \eqref{eq:sn-as}, which gives the survival probability after $t$ generations depending on the maximal number of mutations that can accumulate $n$,

\begin{equation} \nonumber
 S_{1,n}\sim (1+t)^{-\chi_n}+\frac{\chi_n}{2}{(1+t)}^{-\tfrac{1}{2}-\chi_n}
\end{equation}
where $\chi_n=2^{1-n}$.
In Figure \ref{fig:ex1}A, we see that for $n=1$, the survival probability is less than $10\%$ after $10$ generations, whilst up to 100 generations are needed for the same probability if $n=2$. For $n=5$, a reduction in survival probability of $50\%$ will take $10^5$ generations.
In yeast, most experiments determine the growth of colonies after 20 generations. We can compute the expected size of the colony after $t$ generations using equation \eqref{eq:pin}. In Figure \ref{fig:ex1}B, we see that, if only one mutation can accumulate ($n=1$), we won't observe any growth after 20 generations, whilst if two or three mutations can accumulate ($n=2,3$), colonies will form, although smaller than expected. This is qualitatively relatable to the comparison between haploid and diploid yeast investigated by \cite{herr2014dna}, who found that at the limiting mutation rate of one mutation in an essential gene per cell division, haploid yeast colonies are not viable, whilst diploid yeast form viable, but smaller colonies. For $n>4$, the expected growth after 20 generations plateaus to the expected size in a normally exponentially growing population, suggesting that one would need to run longer experiments to observe EEX at the macroscopic level in cell lines that are more robust to mutational burden, in accordance with \ref{fig:ex1}A. The survival probabilities and expected population size for the general case including cell death are available in Appendix \ref{app:ndeath}). Introducing death in the model effectively re-scales time, such that sub-populations with higher death rates go extinct faster.

A remarkable conclusion from the model is that, even if cells are at the limiting mutation rate and hence EEX will occur with probability one, this might take many generations. Thus, in the usual experimental setting, EEX is not detectable by following colony size. However, it would be detectable if sequencing or genetic information of the colonies at different time points was available. In the next Section, we consider the evolution of genetic diversity in more depth.

\subsection{Genetic diversity during EEX}
\begin{figure}
\centering
\includegraphics[scale=0.252]{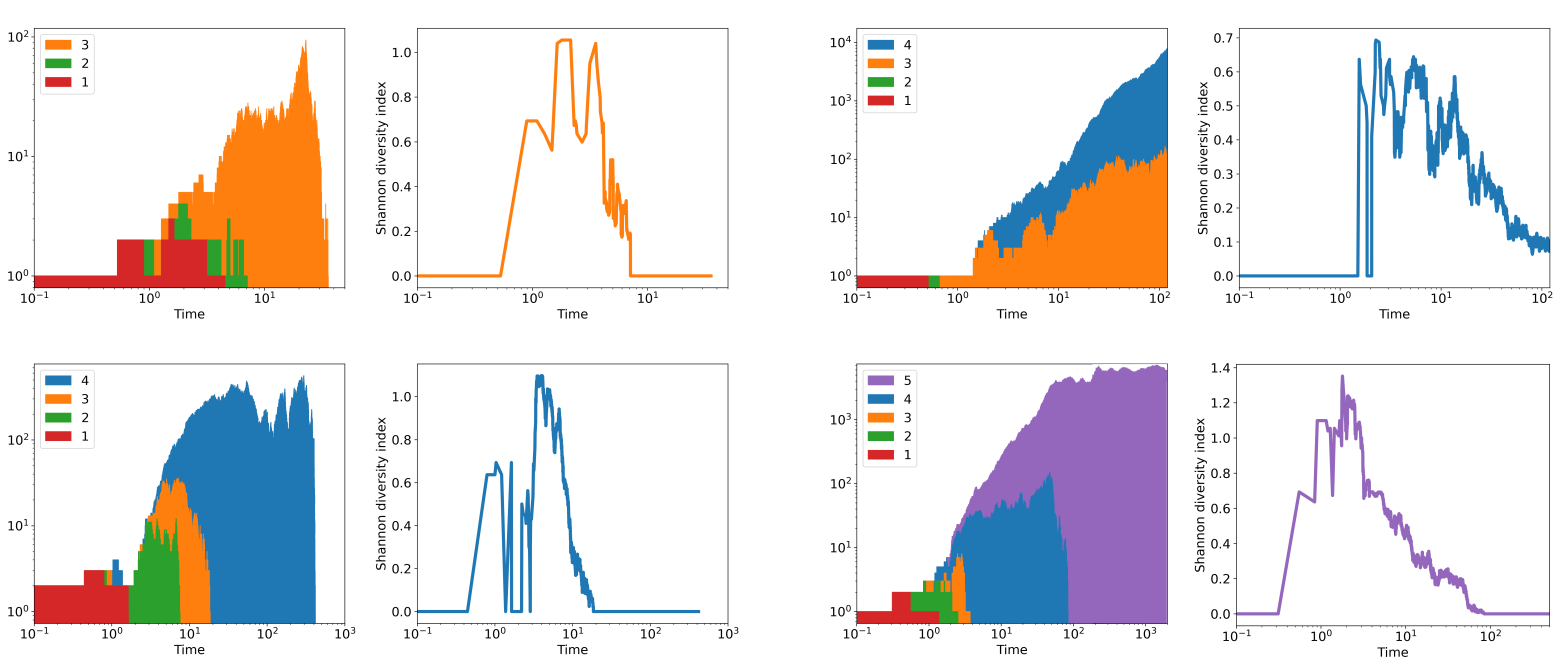} 
\caption{Example simulation runs of the 3-type (A), 4-type (B,C) and 5-type (D) process, where each colored area represents the number of cells of a type, with different types piled on top of each other, and we ran the process until extinction or $10^4$ steps. Next to each run, we plot the corresponding Shannon diversity index $H'(t)$ in time \ref{eq:shannon}. In all examples, this is maximal when the last type arrives, and then quickly decreases as the population becomes dominated by the last type. }
\label{fig:ex2}
\end{figure}

Multi-type branching processes are widely applied to studying the genetic structure of growing populations. In the multi-type critical process, we observe an interesting behaviour, where extreme mutation rates result in an initial increase of genetic diversity, but after a transient phase, the population becomes dominated by cells carrying the maximal number of mutations, and thus loses genetic diversity. In Figure \ref{fig:ex2} we plot the evolution of the number of cells of different simulation runs of the $3$, $4$, and $5$-type process, next to the Shannon diversity index, which is defined as
\begin{equation}
H'(t)=-\sum _{i=1}^{n}p_{i}(t)\log p_i (t)\label{eq:shannon}
\end{equation}
where $p_i(t)$ is the proportion of cells of type-$i$ at time $t$. Indeed, we can see that $H'$ increases as the first mutations accumulate, and rapidly declines when the last type of cells arrive.

\begin{figure}
\centering
\includegraphics[scale=0.25]{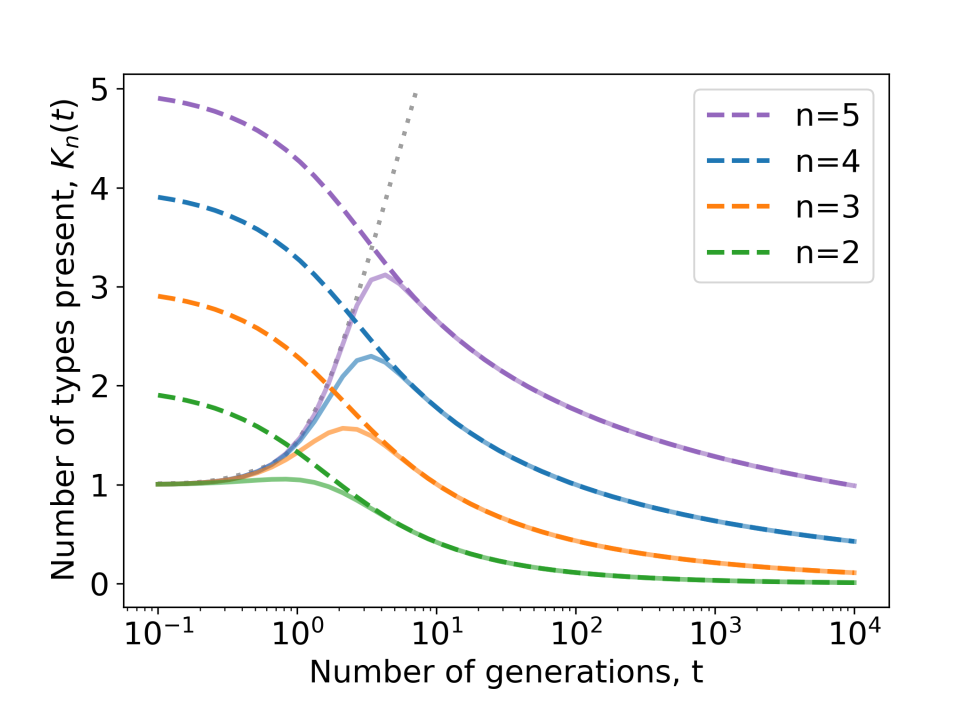} 
\caption{ Mean number of types present at time $t$ as a for populations with different maximal number of mutations ($n$), calculated from \eqref{meantypes} (solid lines) with their large time asymptotic expressions (dashed lines). The dotted grey line shows the expected number of types in a population growing exponentially with no maximal number of mutations ($n=\infty$). }
\label{fig:ex22}
\end{figure}

Even though an analytical expression for $H'(t)$ from the model is not available, we can use \eqref{eq:sn-as} to obtain the expected total number of types or subclones present at time $t$, with its large time asymptotic expression
\begin{equation}
\label{meantypes}
K_n(t):= \mathbb{E}[\#\text{ of types at time } t]= \sum_i^n Q_{1,i}(t) \sim \sum_i^n S_{1,i}(t)
\sim t^{-\chi_n},
\end{equation}
where $Q_{1,i}(t)$ denotes the survival probability of the $i$th cell-type, and $S_{1,i}(t)$ denotes the survival probability of the entire population (defined in \eqref{eq:qn} and \eqref{eq:sndef}, respectively). In Figure \ref{fig:ex22} we see that, similarly to the Shannon diversity index, $K_n(t)$ initially increases and then quickly decreases. As observed in the simulations, after 10-20 generations, the expected number of types present already declines, in sharp contrast with the case in which the colony grows exponentially at a high mutation rate but cells do not acquire lethal mutations. Therefore, even though the macroscopic behaviour is indistinguishable in that timescale, if genetic information on the colonies in time is available experimentally (e.g.\ the number of subclones after $20$ generations), one can identify populations that will eventually reach extinction due to high mutation. Moreover, one can use the expressions derived from the model to infer the maximal number of mutations $n$, which can then be used to quantitatively predict the evolution of the colonies and the time until EEX.

\section{Discussion}

Multi-type branching processes provide a natural tool to model biological processes driven by cell division, death, and mutation. Due to their potential to describe evolutionary dynamics, extensive work has been dedicated to deriving solutions of multi-type processes, especially the super-critical and sub-critical cases \cite{Durrett:2015, nicholson2022mutation}. In this work, we have focused on finite-type critical processes, which mimic populations in which mutations accumulate until a maximal number of alterations is reached, resulting in extinction. 

Driven by biological motivation, we have focused on deriving solutions for the survival of the population, the number of cells, and the arrival and extinction time of cells with different mutations. We found that the survival probability of the overall system, which is asymptotically equivalent to the survival probability of the last type of cells, decays as $t^{-\chi_n}$ for the $n$th type. The $\chi_n = 2^{1-n}$ exponents for the survival probability had been derived by previous work by Foster and Ney \cite{Ney74} and Ogura \cite{ogura1975asymptotic}, although following different approaches. With our approach we have also derived higher-order terms, facilitating the estimation of the accuracy of the first-order term. 

For two cell types, the generating function of the population sizes was expressed explicitly in terms of modified Bessel functions. This provides a numerically efficient way to extract the number distributions, as detailed in Appendix~\ref{app:num}.
 By conditioning on survival of the final type, we extract the distribution of the number of type-$k$ cells in the large time limit. The survival probabilities show us that the system becomes quickly dominated by the last type of cells, and indeed the distribution of the last type coincides with that of the total number of cells. 
 This distribution only depends on the ratio $s/t$ of the size $s$ of the population and time $t$. Interestingly, in the large $s/t$ limit, this has algebraic and stationary tail $(\text{size})^{-1-\chi_n}$. That is, for a fixed (large) number of cells, there is a time regime during which the probability of finding $s$ cells remains constant in time. Since this fat tail -- for each but the first cell type -- would imply infinite population size, we have derived an upper cutoff for the power law tail which ensures the finiteness of the mean population sizes. These power-law tails appearing for large times for $n\ge2$ cell types are in sharp contrast to the purely exponential behaviour of the mass function of the first cell type.
Although the solution is only valid for large time and population size, this corresponds to the range of interest in biological applications.

 We provide exact or asymptotic formulas for our model of EEX in time, including the evolution of population size, and the time of arrival and extinction of sub-populations. In Section \ref{sec:exs}, we discussed example applications of these formulas for studying EEX in microbes, in relation to experimental work. In particular, we showed that populations at the limit mutation rate that can accumulate a high number of mutations take longer to exhibit macroscopic changes in colony size than the number of generations that experiments usually track. Thus, even though such cell lines would eventually undergo error-induced extinction, this extreme fate would be predicted  by analyzing colony growth. In order to identify colonies at the limit mutation rate, experiments should complement macroscopic measures with measures of genetic diversity such as the number of subclones.

Apart from providing tools to quantitatively analyze and guide experimental design, the model provides theoretical insight into error-induced extinction. We propose a general mechanism for EEX: populations of cells that divide and mutate or die at the same rates and have a maximal number of mutations tolerable. In the context of cancer, this maximal number might represent the amount of DNA damage that can accumulate before being detected by the immune system \cite{schumacher2015neoantigens}. 
 We find that the modeled populations become dominated by cells carrying the maximal number of mutations and thus lose genetic diversity. This is related to the idea that genetic instability results in extinction because populations cannot overcome selective barriers \cite{tejero2016theories, andor2017genomic, tilk2022most}. Another interesting behaviour is that populations at the error threshold reach a stationary phase before going extinct. Stationary growth of cancer or bacteria populations is normally associated with having reached a carrying capacity (due to limited nutrients, etc.) \cite{gerlee2013model, tjorve2017use}. Our model shows that this macroscopic behaviour might be caused by mutational burden, in which case the fate of the population is drastically different, highlighting the importance of considering genetic structure when modeling population growth. 
 
A related phenomenon to EEX is the so-called error catastrophe, which describes the inability of a genetic element to be maintained in a population as the fidelity of its replication machinery decreases beyond a certain threshold value, such that it cannot produce enough copies of itself \cite{summers2006examining}. This
has been invoked as a theoretical basis for the treatment of viral infection with drugs that would push the error rate for copying
the viral genome beyond this threshold \cite{vignuzzi2005ribavirin}. In our model, every type of cells eventually disappears due to high mutation, and thus undergoes an error catastrophe. Error catastrophes might result in the most fit cells to be replaced by lower fitness population, but, unlike EEX, not necessarily in extinction. This more general case can be studied using the proposed model, but either allowing for infinite number of mutations, or adjusting the birth and death rate of the last type of cells.  As shown in previous work, the applications of multi-type critical processes extend beyond error catastrophe and EEX, e.g., to modeling infectious disease spread \cite{ak12}.

Throughout this paper, we have focused on the simplest $n$-type critical process, with zero death rate for all but the last cell type. The more general case including cell death, and arbitrary mutation and division rates can be found in Appendix~\ref{app:ndeath}. This could be relevant for  modelling systems in which mutations result in cells with different division, mutation, and death rates, as long as all cell types have critical growth. Including intermediate types with sub-critical and super-critical growth remains a challenge for future work. Future work should consider non-consecutive mutations, allowing for specific mutational paths to be modeled. This is particularly relevant to the study of the fitness of mutator alleles in cancer evolution, which is typically driven by the accumulation of specific mutations.

\appendix

\section{Two types with general rates}
\label{app:two}
The solution for the two-type case has appeared in \cite{Antal:2011}, but we recall the results here for ease of reference with $\alpha_1$ explicitly included in the formulas
\begin{equation}
\label{zgen}
\mathcal{Z}_{1,2} (x_1,x_2,t)= 1-\frac{\nu_1}{\alpha_2 \tau}\,
\frac{I_0(2\tau)-cK_0(2\tau)}{I_1(2\tau) + cK_1(2\tau)}
\end{equation}
where we used the variable
\[\tau = \sqrt{\frac{\nu_1}{\alpha_2} \left(\frac{\alpha_1}{\alpha_2 (1-x_2)} +t \alpha_1 \right)} \]
and the amplitude
\begin{equation}
c = \frac{I_0(2\tau_0)-\alpha_2 \nu_1 ^{-1}\tau_0(1-x_1) I_1(2\tau_0)}
{K_0(2\tau_0) + \alpha_2 \nu_1^{-1}\tau_0 (1-x_1)K_1(2\tau_0)}
\end{equation}
where $\tau_0\equiv \tau(t=0)$. 
\section{Integral formulas $P_{a,b}(t)$}\label{app:if}
The variable $x_1$ appears in \eqref{PA-sol} only through $c$, while $c$ is a ratio of linear (in variable $x_1$) functions. Therefore $\mathcal{Z}_{1,2}$ itself can be transformed into the ratio of linear in $x_2$ functions. We can expand in $x_1$ and effectively we do not need to take the integral in the (complex) $x_1$ plane. Thus for $a=0$ we have 
\begin{equation}
\label{Cauchy:0}
P_{0,b}(t) = \frac{1}{2\pi i}
\oint \frac{d x_2}{ x_2 ^{b+1}}\,\mathcal{Z}_1(0,x_2,t)
\end{equation}
while for $a>0$ 
\begin{equation}
\label{Cauchy:m}
P_{a,b}(t) = \frac{1}{2\pi i}
\oint \frac{d x_2}{ x_2 ^{b+1}}\,\frac{\tau_0^{a}}{\tau}\, 
\frac{(f')^{a-1} (fg'-f' g)}{(f+\tau_0 f')^{a+1}}.
\end{equation}
Here we used shorthand notation
\begin{eqnarray*}
f  &=& I_1(2\tau)K_0(2\tau_0)-K_1(2\tau)I_0(2\tau_0)\\
f' &=& K_1(2\tau)I_1(2\tau_0)-I_1(2\tau)K_1(2\tau_0)\\
g &=& I_0(2\tau)K_0(2\tau_0)-K_0(2\tau)I_0(2\tau_0)\\
g' &=& K_0(2\tau)I_1(2\tau_0)-I_0(2\tau)K_1(2\tau_0).
\end{eqnarray*}
\section{Numerical solutions for $P_{a,b}(t)$ and  $P_s^{(2)}(t)$}\label{app:num}
Explicit solutions for $P_{a,b}(t)$ are not available. Numerically, however, one can access the probabilities. Using Mathematica's \texttt{SeriesCoefficient} command we get 
$$
P_{a,b}(1) \approx \left(\begin{array}{ccccc}
 0.169102 & 0.190391 & 0.0789005 & 0.0341278 &\cdots\\
 0.15659 & 0.0529594 & 0.0224725 & 0.00989238 &\cdots\\
 0.0692559 & 0.0298649 & 0.0138336 & 0.00643762 &\cdots\\
 0.0306302 & 0.0160577 & 0.00810575 & 0.00399019 &\cdots\\
 \vdots & \vdots & \vdots & \vdots &  
\end{array}\right)
$$
for $a,b=0,1,2,3$, which values can be used to check simulations. Simulating the process (for a few seconds on an average laptop with a code in C) over $10^8$ runs we get
$$
P_{a,b}(1) \approx 
\left(
\begin{array}{ccccc}
0.169107 & 0.190410 & 0.078873 & 0.034104 & \cdots\\ 
0.156639 & 0.052987 & 0.022464 & 0.009880 & \cdots\\
0.069277 & 0.029886 & 0.013826 & 0.006427 & \cdots\\
0.030605 & 0.016039 & 0.008102 & 0.003988 & \cdots\\
\vdots & \vdots & \vdots & \vdots &
\end{array}
\right)
$$
However, this approach only works for small $t$. In order to obtain numerical solutions for large $t$, one can invert the generating function via Fourier transform, and apply the Inverse Fast Fourier Transform (IFFT) algorithm as an efficient method to calculate the probability density from the generating function \cite{abate1992fourier}. As an example, we illustrate how to obtain $P_s^{(2)}(t)$, that is, the probability of there being $s$ type-2 cells.
In order to obtain the joint distribution, one needs to invert Equation \eqref{Cauchy:m}. Recall the generating function for type-2 cells 
$$
 \mathcal{Z}_{1,2}(1,x,t)= \sum_{s\geq 0} P^{(2)}_s(t)x^s.
$$
The correspondence to the Fourier transform is easier to see if we consider $z=1/x$,
$$
 \mathcal{Z}_{1,2}(1,1/z,t) = \sum_{s\geq 0} P^{(2)}_s(t)z^{-s}
$$
this has inversion
\begin{equation}
\label{Cauchy:2}
P^{(2)}_s (t)= \frac{1}{2\pi i}
\oint_{C} {dz}\,{z^{s-1}}\,\mathcal{Z}_1(1,1/z,t) 
\end{equation}
where the contour $C$ goes counterclockwise around the origin in the complex $y$ plane, and must enclose all poles of $\mathcal{Z}_1(1,1/z,t)$. We choose $C$ to be a circle of radius $R$ enclosing all poles, and set $z=Re^{i\theta}$,
\begin{align}
P^{(2)}_s (t)&= \frac{R^s}{2\pi }
\int_{0}^{2\pi} {d\theta}\,\mathcal{Z}_1(1,Re^{-i\theta},t){e^{i\theta s}}\label{Cauchy:3}.
\end{align}
We notice that we recover a Fourier series expansion. The above integral can be approximated as a sum by splitting the circle into $N$ equidistant discrete points
\begin{equation}
\label{Cauchy:dis}
{p^{(2)}_s (t)}=\frac{R^s}{N}\sum_{k=0}^{N-1}\mathcal{Z}_1(1,R e^{-2ik\pi /N},t) e^{2isk\pi/N}
\end{equation}
which is the Inverse Discrete Fourier Transform scaled by $ R^{s}$. One can use the IFFT algorithm to obtain the approximate probability density function. The approximation error of ${p^{(2)}_s (t)}\approx P^{(2)}_s (t) $ depends on the discretization \cite{abate1992fourier,cavers1978fast}.
We use the following algorithm to extract the probabilities numerically:
\begin{enumerate}
    \item Calculate $\mathcal{Z}_1(1,R e^{-i\theta},t)=\mathcal{Z}_1(1,1/z,t) , \,z=Re^{i\theta}$.
    \item Discretize the circle into $N$ equidistant points.
    \item Evaluate the generating function at each point. That is, calculate $\mathcal{Z}_1[k]:=\mathcal{Z}_1(1,Re^{- 2 \pi k /N},t)$ for $ k=0,\dots,N-1 $.
    \item Calculate the IDFT of $\mathcal{Z}_1[k]$ using Fast Inverse Fourier Transform (e.g.\ \texttt{fft.ifft} in Numpy or  \texttt{InverseFourierTransform} in Mathematica). This outputs $N$ coefficients $q_0, \dots, q_{N-1}.$
    \item Re-scale the $s$th coefficient by $R^s$ to recover the probability of $s$ cells,
    \[ P^{(2)}_s (t)\approx {p^{(2)}_s (t)}= R^s q_s .\]
\end{enumerate}
Note that the generating function of the two-type system $\mathcal{Z}_{1,2}(x_1,x_2,t)$ has a singularity at $x_2=1$. Thus, we must take $R>1$. From equation \eqref{Cauchy:3} we see that the radius $R$ does not affect the inversion. However, numerical errors due to finite precision number representation result in numerical errors with $R$. In Figure \ref{fig:num} we take $R=1.0001$ as we find that if $R$ is increased it becomes difficult to resolve the separate contributions from the poles and zeros of $\mathcal{Z}_1(1,1/z,t)$. However, in other settings, numerical problems might arise if $R$ is decreased to too close to the furthest pole. In general, it is common practice to take $R$ to be $10\%$ larger than the largest pole \cite{cavers1978fast}. 

\section{Useful formulas and asymptotic results }\label{app:formulas}
We use the large $z$ expansions for the first and second type modified Bessel functions (Section 3.13 of \cite{bender2013advanced}, Section 10.7 of \cite{NIST:DLMF}), 
\begin{align}
I_n(z)&\sim \frac{e^z}{\sqrt{2\pi z}}\left( 1 - \frac{4 n^2 -1}{8z} + \frac{(4 n^2 -1 ) (4 n^2 - 9)}{2! (8z)^2} -\dots \right),
\\ K_n(z)&\sim \sqrt{\frac{\pi}{2z}}e^{-z}\left( 1 - \frac{4 n^2 -1}{8z} + \frac{(4 n^2 -1 ) (4 n^2 - 9)}{2! (8z)^2} - \dots \right).
\end{align}
We often use $I_0(z)/I_1(z)\sim 1$, $K_1(z)\sim 0$ and $K_0(z) \sim 0$. 
On some occasions, we are interested in higher-order terms, which can be obtained by considering more terms in the asymptotic expansions. In particular
$$\frac{I_0(z)}{I_1(z)} = 1+\frac{1}{z}+\frac{1}{2 z^2}+\frac{27}{16 z^3}+\frac{141}{128 z^4}+O\left(z^{-5}\right).$$

\section{$n$ types with general rates}\label{app:ndeath}
Here we generalize our results to the $n$-type process to include death and mutation at arbitrary rates. This is described by the scheme

\begin{equation*}
\begin{split}
\xymatrix{(1)+(1)			& (2)+(2)				&			& (n)+(n)		\\
                (1) \ar[u]_{\alpha_1} \ar[r]_{\nu_1} \ar[d]^{\alpha_1-\nu_1}	& (2)\ar[u]_{\alpha_2} \ar[r]_{\nu_2} \ar[d]^{\alpha_2-\nu_2} 	& \cdots \ar[r]_{\nu_{n-1}} 	& (n) \ar[u]_{\alpha_n} \ar[d]^{\alpha_n}\\
                \emptyset			& \emptyset			& 			& \emptyset }
\end{split}                
\end{equation*}
Let $\mathcal{Z}_{i,n}(x_1,x_2,\dots x_n, t)$ denote the generating function starting with a single initial type-$i$ cell. The backward Kolmogorov equations read
\begin{align*}
 \partial_t \mathcal{Z}_{i,n} =\begin{cases} \alpha_i \mathcal{Z}_{i,n}^2+ \alpha_i-\nu_i + \nu_i\mathcal{Z}_{i+1,n}-2\alpha_i \mathcal{Z}_{i,n} & \text{for}\quad {i=1,\dots,n-1}\\ 
 \alpha_n (\mathcal{Z}_{n,n}^2 + 1 - 2\mathcal{Z}_{n,n}) &\text{for}\quad {i=n}. \end{cases}
\end{align*}

\subsection{Survival probabilities}
We solve the system for the survival probabilities $S_{i,n}(t)=1-\mathcal{Z}_{i,n}(0,0,\dots,0,t)$, given by
\begin{align*}
 \frac{d S_{i,n}}{dt} =\begin{cases}\nu_i S_{i+1,n}- \alpha_i {S_{i,n}}^2 \quad&\text{for}\quad{i=1,\dots,n-1},\\
   - \alpha_n S_{n,n}^2 \quad&\text{for}\quad{i=n},\end{cases}
\end{align*}
with initial conditions $S_{i,n}(0)=1$ for all $i=1,\dots,n$. Following the procedure outlined in Section~\ref{sec:n-survival} for the no-death case, we derive the survival probability of the $n$-type process
\[S_{1,n}(t)\sim{\prod_{i=1}^{n-1} {\left(\frac{\nu_{i}}{\alpha_{i}} \right)}^{\chi_{i+1}}(\alpha_n t +1)^{-\chi_n}}\]
where $\chi_n=2^{1-n}$. Note that this system is equivalent to the multi-type  critical process without death, up to a re-scaling of time. The asymptotic solutions are exact to the same order, but the convergence is slower. 

\subsection{Generating functions and total number distribution}
We now turn to the generating functions. Again, we already know that
\[\mathcal{Z}_{n,n}=1-\frac{1}{\alpha_n t   + {( 1-x_n)}^{-1} }\]
and we know that for $t\to \infty$, in the leading order $\mathcal{Z}_{i,n}$ only depends on the last type $x_n$, and $x_n \to 1$ so we may assume that in the leading order
\[1- \mathcal{Z}_{i,n} \sim (B_i( t +  {( 1-x_n)}^{-1}) )^{-A_i}\]
and match coefficients as usual to recover the coefficients $B_i$ and $A_i$ for all types. We arrive at the following expression for the leading order asymptotic generating function starting with a single type-1 cell,
\[1-\mathcal{Z}_{1,n}(x_1,\dots, x_n, t)\sim  {\prod_{i=1}^{n-1} {\left(\frac{\nu_{i}}{\alpha_{i}} \right)}^{\chi_{i+1}} (\alpha_n t + {(1-x_n)}^{-1})^{-\chi_n}}.\]
Now we follow the procedure of the no-death case outlined in Section~\ref{sec:n-gener}, use the scaling variables $x_n=1-p/\alpha_n t$ and $y=s/\alpha_n t$ and take the limit $t\to \infty$ with $p$ and $y$ constants, to find 
\begin{equation*}
\mathbb{E}(x_n^{Z_n(t)}|Z_n(t)>0) \to \int_0^\infty f_{Y_n}(y) e^{-py} dy 
= 1-{\left(\frac{p}{p+1}\right)}^{\chi_n}.
\end{equation*}
where we established the convergence in distribution
$$
t\frac{P^{(n)}_s(t)}{Q_{1,n}(t)} \to f_{Y_n}(y)
, \quad \frac{Z_n(t)}{t} | \{Z_n(t)>0\} \to Y_n
$$
Hence we find that the limit random variable $Y_n$ does not depend on the parameters of the system, and has the density \eqref{Pzn}.
Therefore, the scaling form of the $n$-type distribution is
\begin{equation}
\begin{split}
P^{(n)}_{s}(t)&\sim \chi_n \frac{Q_{1,n}(t)}{t \alpha_n} F(1 + \chi_n,2,-s/t\alpha_n)  \\
& \sim {\prod_{i=1}^{n-1} {\left(\frac{\alpha_{i}}{\nu_{i}} \right)}^{\chi_{i+1}} } {\left(\alpha_n t \right)}^{\chi_n -1}\chi_n  F(1 + \chi_n,2,-s/t\alpha_n).\end{split}\end{equation}
Finally, by taking the large argument asymptotic of the confluent hypergeometric function (Appendix \ref{app:formulas}), we obtain the stationary distribution

\begin{equation}
\label{tail:gn}
P^{(n)}_{s}(t) \approx \frac{\chi_n \alpha_n^{\chi_n -1} {\prod_{i=1}^{n-1} {\left(\frac{\alpha_{i}}{\nu_{i}}  \right)}^{\chi_{i+1}}  } }{\Gamma(1-\chi_n)}\,s^{-1-\chi_n}.
\end{equation}
As seen before, the algebraic stationary tail describes the behaviour up to an upper-cutoff $s^{\star}$ in the number of cells, which we recover by matching the order of magnitude of the mean $\mathbb{E} Z_n(t)$,
\begin{equation}
 t\ll s\ll t^{\tfrac{n-1}{1-\chi_n}}.
\end{equation}


\subsection{Arrival times} We now consider the arrival times for different birth and mutation rates for each type, $\alpha_i$, $\nu_i$ (as described by the scheme \eqref{fig:bd_multid}). The equations for $g_{i,j}$ satisfy
\begin{equation}\frac{d  g_{i,j}}{dt}  = - \alpha_i g_{i,j}^2 + \nu_i g_{i+1,j}\label{eq:gij death gen}
\end{equation} 
with initial conditions $g_{i,j}(0)=0$ and $g_{j,j}\equiv 1$.
This leads to the asymptotic probability of arrival of the $j$th type starting with a single cell
\begin{equation}\label{ginf}
    g_{1,j}(\infty):=\lim_{t\to \infty} g_{1,j}(t)=\prod_{i=1}^{j-1} {\left(\tfrac{\nu_i}{\alpha_i}\right)}^{2^{-i}} .
\end{equation}
As in the pure birth-mutation case \eqref{smallnu}, we obtain a small mutation rate limit (all $\nu_i\to 0$, starting with $\nu_{j-1}$) by normalizing and re-scaling time by \eqref{ginf}
$$
g_{1,j}\left(t \right)\approx g_{1,j}(\infty) \tanh{g_{1,j}(\infty) t}.
$$

\bibliographystyle{plain}
\bibliography{biblio}

\begin{thebibliography}{10}

\bibitem{abate1992fourier}
Joseph Abate and Ward Whitt.
\newblock The fourier-series method for inverting transforms of probability
  distributions.
\newblock {\em Queueing systems}, 10:5--87, 1992.

\bibitem{albertson2009dna}
Tina~M Albertson, Masanori Ogawa, James~M Bugni, Laura~E Hays, Yang Chen,
  Yanping Wang, Piper~M Treuting, John~A Heddle, Robert~E Goldsby, and
  Bradley~D Preston.
\newblock Dna polymerase $\varepsilon$ and $\delta$ proofreading suppress
  discrete mutator and cancer phenotypes in mice.
\newblock {\em Proceedings of the National Academy of Sciences},
  106(40):17101--17104, 2009.

\bibitem{andor2017genomic}
Noemi Andor, Carlo~C Maley, and Hanlee~P Ji.
\newblock Genomic instability in cancer: teetering on the limit of tolerance.
\newblock {\em Cancer research}, 77(9):2179--2185, 2017.

\bibitem{Antal:2010}
Tibor Antal and P~L Krapivsky.
\newblock Exact solution of a two-type branching process: clone size
  distribution in cell division kinetics.
\newblock {\em Journal of Statistical Mechanics: Theory and Experiment},
  2010(07):P07028, 2010.

\bibitem{Antal:2011}
Tibor Antal and P~L Krapivsky.
\newblock Exact solution of a two-type branching process: models of tumor
  progression.
\newblock {\em Journal of Statistical Mechanics: Theory and Experiment},
  2011(08):P08018, 2011.

\bibitem{ak12}
Tibor Antal and P~L Krapivsky.
\newblock Outbreak size distributions in epidemics with multiple stages.
\newblock {\em Journal of Statistical Mechanics: Theory and Experiment},
  2012(07):P07018, 2012.

\bibitem{Athreya:2004}
K~B Athreya and P~E Ney.
\newblock {\em {Branching Processes}}.
\newblock Dover Publications, 2004.

\bibitem{bender2013advanced}
Carl~M Bender and Steven~A Orszag.
\newblock {\em Advanced mathematical methods for scientists and engineers I:
  Asymptotic methods and perturbation theory}.
\newblock Springer Science \& Business Media, 2013.

\bibitem{cavers1978fast}
JK~Cavers.
\newblock On the fast fourier transform inversion of probability generating
  functions.
\newblock {\em IMA Journal of Applied Mathematics}, 22(3):275--282, 1978.

\bibitem{Chistyakov59}
V~P Chistyakov.
\newblock Generalization of a theorem for branching processes.
\newblock {\em Theory of Probability \& Its Applications}, 4(1):103--106, 1959.

\bibitem{NIST:DLMF}
{\it NIST Digital Library of Mathematical Functions}.
\newblock \url{https://dlmf.nist.gov/}, Release 1.1.10 of 2023-06-15.
\newblock F.~W.~J. Olver, A.~B. {Olde Daalhuis}, D.~W. Lozier, B.~I. Schneider,
  R.~F. Boisvert, C.~W. Clark, B.~R. Miller, B.~V. Saunders, H.~S. Cohl, and
  M.~A. McClain, eds.

\bibitem{Durrett:2015}
Rick Durrett.
\newblock {\em {Branching Process Models of Cancer}}.
\newblock Stochastics in Biological Systems. Springer, 2015.

\bibitem{fijalkowska1996mutants}
Iwona~J Fijalkowska and Roel~M Schaaper.
\newblock Mutants in the exo i motif of escherichia coli dnaq: defective
  proofreading and inviability due to error catastrophe.
\newblock {\em Proceedings of the National Academy of Sciences},
  93(7):2856--2861, 1996.

\bibitem{Ney74}
James Foster and Peter Ney.
\newblock Decomposable critical multi-type branching processes.
\newblock {\em Sankhyā: The Indian Journal of Statistics, Series A
  (1961-2002)}, 38(1):28--37, 1976.

\bibitem{foster1978limit}
James Foster and Peter Ney.
\newblock Limit laws for decomposable critical branching processes.
\newblock {\em Zeitschrift f{\"u}r Wahrscheinlichkeitstheorie und Verwandte
  Gebiete}, 46(1):13--43, 1978.

\bibitem{fox2010lethal}
Edward~J Fox and Lawrence~A Loeb.
\newblock Lethal mutagenesis: targeting the mutator phenotype in cancer.
\newblock In {\em Seminars in cancer biology}, volume~20, pages 353--359.
  Elsevier, 2010.

\bibitem{gerlee2013model}
Philip Gerlee.
\newblock The model muddle: in search of tumor growth laws.
\newblock {\em Cancer research}, 73(8):2407--2411, 2013.

\bibitem{herr2014dna}
Alan~J Herr, Scott~R Kennedy, Gary~M Knowels, Eric~M Schultz, and Bradley~D
  Preston.
\newblock Dna replication error-induced extinction of diploid yeast.
\newblock {\em Genetics}, 196(3):677--691, 2014.

\bibitem{herr2011mutator}
Alan~J Herr, Masanori Ogawa, Nicole~A Lawrence, Lindsey~N Williams, Julie~M
  Eggington, Mallika Singh, Robert~A Smith, and Bradley~D Preston.
\newblock Mutator suppression and escape from replication error--induced
  extinction in yeast.
\newblock {\em PLoS genetics}, 7(10):e1002282, 2011.

\bibitem{kesten67}
H~Kesten and B.P Stigum.
\newblock Limit theorems for decomposable multi-dimensional galton-watson
  processes.
\newblock {\em Journal of Mathematical Analysis and Applications},
  17(2):309--338, 1967.

\bibitem{morrison1993pathway}
Alan Morrison, Anthony~L Johnson, Leland~H Johnston, and Akio Sugino.
\newblock Pathway correcting dna replication errors in saccharomyces
  cerevisiae.
\newblock {\em The EMBO journal}, 12(4):1467--1473, 1993.

\bibitem{mullikin1963limiting}
Thomas~W Mullikin.
\newblock Limiting distributions for critical multitype branching processes
  with discrete time.
\newblock {\em Transactions of the American Mathematical Society},
  106(3):469--494, 1963.

\bibitem{nicholson2022mutation}
Michael~D Nicholson, David Cheek, and Tibor Antal.
\newblock Mutation accumulation in exponentially growing populations.
\newblock {\em arXiv preprint arXiv:2208.02088}, 2022.

\bibitem{ogura1975asymptotic}
Yukio Ogura.
\newblock Asymptotic behavior of multitype galton-watson processes.
\newblock {\em J. Math. Kyoto Univ}, 15(2):251--302, 1975.

\bibitem{schumacher2019cancer}
Ton~N Schumacher, Wouter Scheper, and Pia Kvistborg.
\newblock Cancer neoantigens.
\newblock {\em Annual review of immunology}, 37:173--200, 2019.

\bibitem{schumacher2015neoantigens}
Ton~N Schumacher and Robert~D Schreiber.
\newblock Neoantigens in cancer immunotherapy.
\newblock {\em Science}, 348(6230):69--74, 2015.

\bibitem{sevast1959transient}
Boris~Alexandrovich Sevast’yanov.
\newblock Transient phenomena in branching stochastic processes.
\newblock {\em Theory of Probability \& Its Applications}, 4(2):113--128, 1959.

\bibitem{soriano2021expression}
Ignacio Soriano, Enrique Vazquez, Nagore De~Leon, Sibyl Bertrand, Ellen
  Heitzer, Sophia Toumazou, Zhihan Bo, Claire Palles, Chen-Chun Pai, Timothy~C
  Humphrey, et~al.
\newblock Expression of the cancer-associated dna polymerase $\varepsilon$
  p286r in fission yeast leads to translesion synthesis polymerase dependent
  hypermutation and defective dna replication.
\newblock {\em PLoS genetics}, 17(7):e1009526, 2021.

\bibitem{summers2006examining}
Jesse Summers and Samuel Litwin.
\newblock Examining the theory of error catastrophe.
\newblock {\em Journal of virology}, 80(1):20--26, 2006.

\bibitem{tejero2016theories}
H{\'e}ctor Tejero, Francisco Montero, and Juan~Carlos Nu{\~n}o.
\newblock Theories of lethal mutagenesis: from error catastrophe to lethal
  defection.
\newblock {\em Quasispecies: From Theory to Experimental Systems}, pages
  161--179, 2016.

\bibitem{tilk2022most}
Susanne Tilk, Svyatoslav Tkachenko, Christina Curtis, Dmitri~A Petrov, and
  Christopher~D McFarland.
\newblock Most cancers carry a substantial deleterious load due to
  hill-robertson interference.
\newblock {\em Elife}, 11:e67790, 2022.

\bibitem{tjorve2017use}
Kathleen~MC Tj{\o}rve and Even Tj{\o}rve.
\newblock The use of gompertz models in growth analyses, and new gompertz-model
  approach: An addition to the unified-richards family.
\newblock {\em PloS one}, 12(6):e0178691, 2017.

\bibitem{topatana2020advances}
Win Topatana, Sarun Juengpanich, Shijie Li, Jiasheng Cao, Jiahao Hu, Jiyoung
  Lee, Kenneth Suliyanto, Diana Ma, Bin Zhang, Mingyu Chen, et~al.
\newblock Advances in synthetic lethality for cancer therapy: cellular
  mechanism and clinical translation.
\newblock {\em Journal of hematology \& oncology}, 13:1--22, 2020.

\bibitem{vignuzzi2005ribavirin}
Marco Vignuzzi, Jeffrey~K Stone, and Raul Andino.
\newblock Ribavirin and lethal mutagenesis of poliovirus: molecular mechanisms,
  resistance and biological implications.
\newblock {\em Virus research}, 107(2):173--181, 2005.

\end{thebibliography}
\end{document}